\documentclass[12pt]{elsarticle}

\usepackage{caption}
\usepackage{subcaption}

\usepackage[cmex10]{amsmath,mathtools}
\usepackage{mdwmath}
\usepackage{mdwtab}
\usepackage{amssymb}

\usepackage{algorithmic}
\usepackage{algorithm}
\usepackage{picins}
\usepackage{float}

\usepackage{epstopdf}
\usepackage{multirow}
\usepackage{color}

\begin{document}

\begin{frontmatter}

\title{On-board Deep Q-Network for UAV-assisted Online Power Transfer and Data Collection}

\author[cister]{Kai Li}
\ead{kaili@isep.ipp.pt}

\author[csiro]{Wei~Ni}

\author[cister]{Eduardo~Tovar}

\address[cister]{Real-Time and Embedded Computing Systems Research Centre, Portugal}

\address[csiro]{Commonwealth Scientific and Industrial Research Organization, Australia}

\begin{abstract}
Unmanned Aerial Vehicles (UAVs) with Microwave Power Transfer (MPT) capability provide a practical means to deploy a large number of wireless powered sensing devices into areas with no access to persistent power supplies. The UAV can charge the sensing devices remotely and harvest their data. A key challenge is online MPT and data collection in the presence of on-board control of a UAV (e.g., patrolling velocity) for preventing battery drainage and data queue overflow of the sensing devices, while up-to-date knowledge on battery level and data queue of the devices is not available at the UAV. In this paper, an on-board deep Q-network is developed to minimize the overall data packet loss of the sensing devices, by optimally deciding the device to be charged and interrogated for data collection, and the instantaneous patrolling velocity of the UAV. Specifically, we formulate a Markov Decision Process (MDP) with the states of battery level and data queue length of sensing devices, channel conditions, and waypoints given the trajectory of the UAV; and solve it optimally with Q-learning. Furthermore, we propose the on-board deep Q-network that can enlarge the state space of the MDP, and a deep reinforcement learning based scheduling algorithm that asymptotically derives the optimal solution online, even when the UAV has only outdated knowledge on the MDP states. Numerical results demonstrate that the proposed deep reinforcement learning algorithm reduces the packet loss by at least 69.2\%, as compared to existing non-learning greedy algorithms.
\end{abstract}


\end{frontmatter}

%=============================================================================%
%============================ Section 1 Introduction================================%
\section{Introduction}
\label{sec_intro}
\begin{figure*}[htb]
\centering
\includegraphics[width=5in]{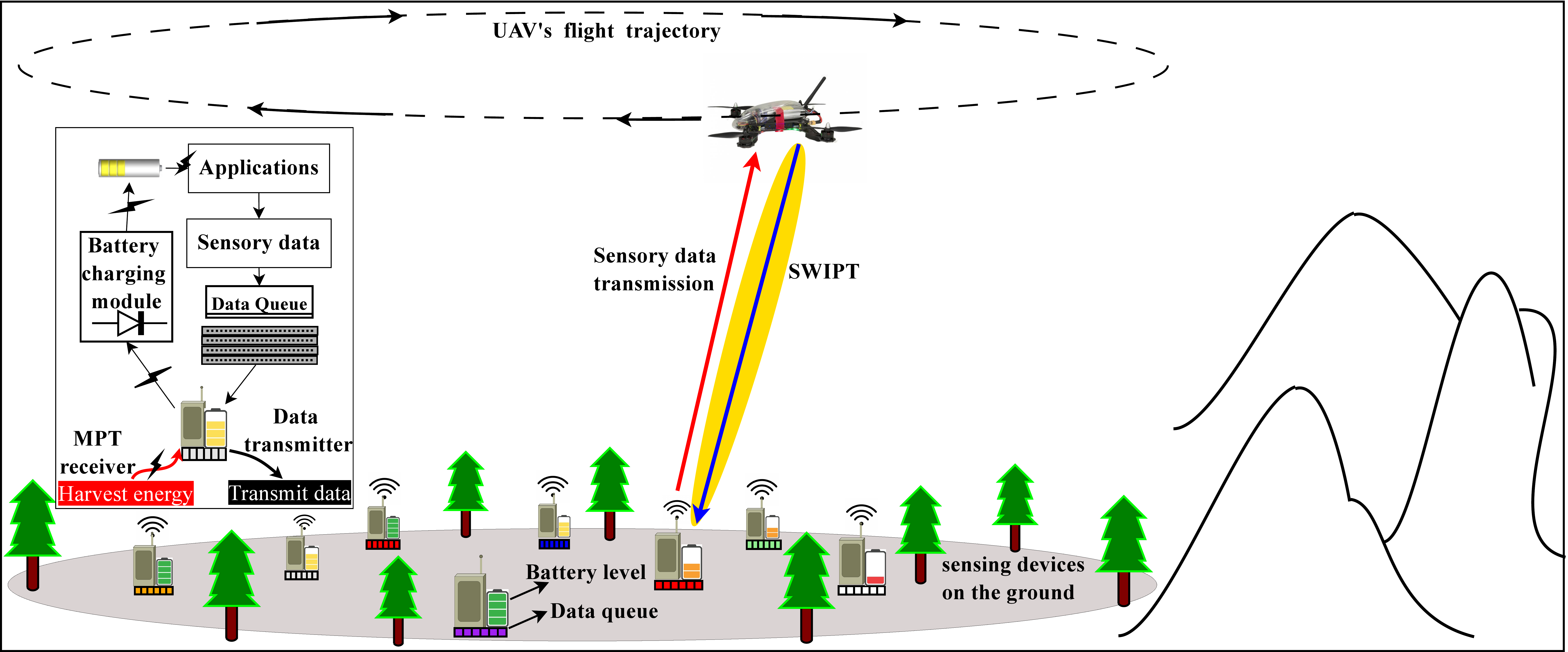}
\caption{\small{UAV-assisted MPT and data collection with the wireless powered ground devices.} }
\label{fig_network}
\end{figure*}

Wireless sensing devices operating on limited batteries have been airlifted and deployed in remote, human-unfriendly environments. 
Conventional terrestrial communication networks and persistent power supplies are unavailable or unreliable in such harsh environments. 
Unmanned Aerial Vehicles (UAVs) have been proposed to harvest data from the sensing devices, and offload command or software pitch, thanks to UAVs' excellent mobility and maneuverability, flexible deployment, and low operational costs~\cite{tomic2012toward,luo2012communication,waharte2010supporting}. 
Figure~\ref{fig_network} depicts a typical real-time application using the UAV in rescue operations, where the wireless sensing device on the ground, typically running with limited battery, records critical information, such as locations, temperature, health conditions of rescuers, and oxygen supply. Each ground device generates data packets at an application-specific sampling rate, and put them into a data queue for future transmission. 

A UAV can be employed to hover over the area of interest, collecting and ferrying the sensory data of the ground device. 
Since the ground device is constrained by a typically limited battery power which limits scalability and sustainability of the sensor network, Microwave Power Transfer (MPT) has been studied to enable energy harvesting in UAV-assisted data collection~\cite{wang2018power,zeng2016wireless}. 
Particularly, we consider a power-splitting MPT technique, e.g., Simultaneous Wireless Information and Power Transfer (SWIPT), where the UAV sends response messages to the ground device meanwhile charging its batteries~\cite{perera2017simultaneous,yin2017uav,yin2018uav}. 
Moreover, the ground device, equipped with a data communication antenna and a wireless power receiver, collects electrical energy conveyed by radio frequency signals while recovering the data from the signals. With SWIPT, every ground device only needs a single RF chain with a reduced hardware cost, since MPT and data transmission can work in the same radio frequency band. 

In practical scenarios, energy harvesting and data transmission could be severely affected by movements of the UAV and time-varying channels. Moreover, the up-to-date knowledge about battery level and data queue length of the ground devices is not available at the UAV. 
Therefore, scheduling MPT and data collection online in the presence of on-board control of the UAV (e.g., patrolling velocity) for preventing battery drainage and data queue overflow is critical in UAV-assisted wireless powered sensor networks. 

In this paper, the problem is formulated as a Markov Decision Process (MDP) with the states of battery level and data queue length of the ground devices, channel conditions, and waypoints given the trajectory of the UAV, which can be optimally solved by a reinforcement learning approach, e.g., Q-learning. However, Q-learning suffers from the well-known curse of dimensionality, which is impractical for the resource allocation in the UAV-assisted online MPT and data collection due to a large number of states and actions. 
Furthermore, we propose an on-board deep Q-network that can enlarge the state and action space of the MDP to minimize the data packet loss of the entire system. 
A new Deep Reinforcement Learning based Scheduling Algorithm (DRL-SA) is developed, which derives the optimal solution online by taking current network state and action as the input and delivering a corresponding action-value function. 
DRL-SA learns an optimal resource allocation strategy asymptotically through online training at the on-board deep Q-network, where the selection of the ground device, modulation scheme, and instantaneous patrolling velocity of the UAV are jointly optimized based on the action-value function. 
DRL-SA utilizes an $\epsilon$-greedy policy to balance the network cost minimization with respect to the knowledge already known with trying new actions to obtain knowledge unknown. 
Moreover, DRL-SA carries out experience replay~\cite{lin1993reinforcement} to significantly reduce expansion of the state space in which the algorithm's scheduling experiences at each time step are stored in a data set. 
DRL-SA is implemented using Keras deep learning library with Google TensorFlow as the backend engine. 
Numerical results demonstrate that the proposed DRL-SA is able to reduce the packet loss by 69.2\%, as compared to existing non-learning greedy algorithms.

In our earlier work~\cite{li2018reinforcement,li2018wireless}, scheduling strategies were studied with a focus on reducing packet loss in small-scale static wireless powered sensor networks in response to battery level and data queue statues of the ground devices. Due to the low-dimensional channel and device state spaces, the resource allocation problem can be solved by reinforcement learning or dynamic programming. 
However, in the UAV-assisted MPT and data collection, the mobility of the UAV with the varying patrolling velocity causes rapidly changing wireless channels. As a result, both the state space and the action space are exceedingly large and grow dramatically fast with the size of the network. This prevents conventional resource allocation approaches, such as~\cite{li2018reinforcement} and~\cite{li2018wireless}, from scaling to the high-dimensional input spaces. 

The rest of this paper is organized as follows. 
Section~\ref{sec_relatedwork} presents related work on UAV-assisted data communication with power transfer. 
Section~\ref{sec_systemprotocol} studies system structure of UAV-assisted MPT and data collection, as well as the communication protocol design. 
In Section~\ref{sec_DRL}, a deep reinforcement learning algorithm is proposed to address the resource allocation problem for UAV-assisted online MPT and data collection. 
Numerical results and evaluation are presented in Section~\ref{sec_evaluation}. 
Section~\ref{sec_cond} concludes the paper.

%=============================================================================%
%============================Section 2 Related Work===============================%
\section{Related Work}
\label{sec_relatedwork}
In this section, we review the literature on data communication, trajectory planning, and power transfer in UAV networks. 

\subsection{UAV-assisted wireless communications} 
An energy-efficient UAV relaying scheme is studied in~\cite{li2016energy} to schedule the data transmission between the UAV and the sensor nodes while guaranteeing packet success rate. A practical and computationally efficient algorithm is designed to extend network lifetime, by decoupling energy balancing and modulation adaptation of the UAVs, and optimizing in an alternating manner. 
The authors in~\cite{koulali2016green} implement passive scanning at the sensor nodes and periodic beaconing at the UAV to reduce network energy consumption. A non-cooperative game is constructed and equilibrium beaconing period durations are characterized for the UAVs. A learning algorithm is described to allow a UAV to discover the equilibrium beaconing strategy without observing the other UAVs' relaying schedules. 
Network outage probability is analyzed in which a number of UAVs are used to relay the source signals to a data sink in a decode-and-forward manner~\cite{li2010multi}. The use of multiple antennas at the data sink is considered and their respective performance is also examined. 
The network outage problem can be decoupled into power allocation and trajectory planning subproblems~\cite{zhang2018joint}. An approximate solution that iteratively addresses the two subproblems is developed for approaching the minimum outage probability. In each iteration, the trajectory is planned according to the power control results obtained by the last iteration, and then the power control subproblem is solved given the UAV trajectory. 
A resource allocation algorithm is presented for improving network throughput while guaranteeing a seamless relaying transmission to the data sink outside network coverage via UAV~\cite{baek2018optimal}. By analyzing the outage probability, it is shown that non-orthogonal transmissions improve the performance of the UAV relaying network over orthogonal transmissions. 

\subsection{UAV trajectory planning} 
UAV trajectory planning is studied in~\cite{fadlullah2016dynamic,choi2014energy,jiang2012optimization,zhan2011wireless}, where the UAV acts as a communication relay for connecting the sensor nodes. Network throughput and communication delay can be improved by the motion control of the UAV, e.g., heading, velocity, or radius of the flight trajectory. 
In~\cite{wu2018joint,zeng2016throughput}, the transmit power of the UAV and the trajectory planning are studied to increase network throughput over a finite time horizon. The power allocation exploits the predictable channel changes induced by the UAV's movement while the trajectory planning balances the throughput between the source-UAV and UAV-sink links. 
The UAV can also be used as a buffer-aided relay in free-space optical systems, which stores data emanating from some stationary sensor nodes for possible future delivery to other nodes~\cite{fawaz2018uav}. Since an optical link between the two transceivers can be easily smeared by ambient atmospheric conditions, e.g., an intervening cloud, the UAV adjusts its altitude in such a way that would eliminate the cloud attenuation effect. 

The existing resource allocation approaches in the UAV relaying network improve the performance based on power control and trajectory planning of the UAV. However, scheduling energy harvesting and data collection was yet to be considered. 

\subsection{UAV-assisted power transfer} 
Several studies have integrated MPT technologies into the UAV network. 
The UAV carrying an MPT transmitter is used to collect sensory data and extend the lifetime of sensor networks in harsh terrains~\cite{pang2014efficient,johnson2013charge}. With consideration of transferred power attenuation, residual energy and buffered data at the sensor node, the UAV is scheduled to selectively charge the nodes and collect their data. 
In~\cite{xu2018uav}, the UAV trajectory planning with the velocity control is exploited for charging all the sensor nodes in a fair fashion. The problem formulation and solution imply that the hovering location and duration can be designed to enhance the MPT efficiency. 
Moreover, machine learning techniques can be utilized to predict the UAV's trajectory and improve the energy harvesting efficiency~\cite{jeong2017design}. 
In~\cite{yin2018uav}, SWIPT is introduced into the UAV relaying network, where the sensor node's energy limit can be alleviated by scavenging wireless energy from radio signals transmitted by the UAV. The network throughput is improved by adjusting the UAV's transmit power and the flight trajectory. 
Several MPT platforms are developed for the UAV to charge batteries of the sensor nodes remotely from the electric grid~\cite{he2017drone,wang2016design,chen2016mobile,mittleider2016experimental,griffin2012resonant}. The lightweight design of hardware, control algorithms, and experiments are presented to verify the feasibility, reliability and efficiency of the UAV-assisted power transfer. 
However, the existing literature only focuses on improving the energy efficiency of MPT. The data loss caused by buffer overflows and poor channels is not considered.

%=============================================================================%
%============================Section 3 System Model===============================%
\section{System Model and Communication Protocol}
\label{sec_systemprotocol}
In this section, we introduce the system model and the communication protocol of UAV-assisted online MPT and data collection. Notations used in the paper are summarized in Table~\ref{tb_variables}. 

\subsection{System model}
\label{sec_system}
\begin{table}[htb]
    \centering
    \caption{The list of fundamental variables defined in system model}
    \begin{tabular} {|p{2.8cm}|p{8.5cm}|} \hline
        \bf{Notation} & \bf{Definition} \\ \hline
        	$I$ &  number of wireless powered ground devices  \\[1pt] \hline
	$v_{\rm max}$, $v_{\rm min}$ &  the maximum and minimum velocity of the UAV  \\[1pt] \hline 
	$Z$ &  number of laps the UAV patrols \\[1pt] \hline 
	$P^z_i(t)$ &  transmit power of device $i$ \\[1pt] \hline
	$\tilde{P}^z_i(t)$ &  power transferred to device $i$ \\[1pt] \hline
	$P_{\rm UAV}^{tx}$ &  transmit power of the UAV \\[1pt] \hline
	$\zeta_z(t)$ &  location of the UAV on its trajectory \\[1pt] \hline
	$\mathbf{h}_i^z(t)$ &  channel gain between device $i$ and the UAV  \\[1pt] \hline
	$q_{i,z}(t)$ &  queue length of device $i$ \\[1pt] \hline
	$D$ &  maximum queue length of the ground device \\[1pt] \hline
	$\phi_i^z(t)$ &  modulation scheme of device $i$ \\[1pt] \hline
	$\Phi$ &  the highest modulation order \\[1pt] \hline
	$\gamma_i$ &  SNR between device $i$ and the UAV  \\[1pt] \hline
	$\omega(d_i^z(t),\theta_i^z(t))$ &  MPT efficiency factor  \\[1pt] \hline
	$e_{i,z}(t)$ &  battery level of device $i$ \\[1pt] \hline
	$E$ &  battery capacity of the ground device  \\[1pt] \hline
	$K$ &  the highest battery level of the ground device  \\[1pt] \hline
	$B$ &  number of bits of the data packet \\[1pt] \hline
	$\varepsilon$ &  required BER of the channel between the ground device and the UAV \\[1pt] \hline
	$\widehat{T}^z_i(t)$ & contact time between device $i$ and the UAV \\[1pt] \hline
	$\mathcal{A}$ & action set of MDP \\[1pt] \hline
	$\delta$ & discount factor for future states \\[1pt] \hline
	$l$ & learning iteration in the deep Q-network \\[1pt] \hline
	$\varphi_l$ & learning weight in the deep Q-network \\[1pt] \hline
   \end{tabular}
\label{tb_variables}
\end{table}
\vspace{-2pt}

The network that we consider consists of $I$ wireless powered ground devices in a remote area. The UAV that acts as a data collection node flies a predetermined circular trajectory for $Z$ laps, where the number of laps is limited by the lifetime of the UAV's battery. 
Let $v_{\rm max}$ and $v_{\rm min}$ denote the maximum and minimum patrolling velocity of the UAV, respectively. The patrolling velocity at time slot $t$, i.e., $v^z(t)$, can be adjusted during the flight, where $v^z(t) \in [v_{\rm min}, v_{\rm max}]$ and $z \in [1, Z]$. 
The location of the UAV on its trajectory at $t$ in lap $z$ is denoted by $\zeta_z(t)$. 
The UAV is also responsible for remotely charging the ground devices using MPT. Specifically, receive beamforming is enabled at the UAV to enhance the received signal strength (RSS) and reduce bit error rate (BER). Device $i$ ($ \in [1, I]$) harvests energy from the UAV to power its operations, e.g., sensing, computing and communication. 
In addition, multi-user beamforming techniques, e.g., zero-forcing beamforming, conjugate beamforming, and singular value decomposition, can be applied to UAV-assisted MPT and data collection. However, they are not considered in this work, due to their requirement of real-time feedback on channel state information. 

The complex coefficient of the reciprocal wireless channel between the UAV and device $i$ at $t$ in lap $z$ is $\mathbf{h}_i^z(t)$, which can be known by channel reciprocity. 
The modulation scheme of device $i$ at $t$ in lap $z$ is denoted by $\phi_i^z(t)$. In particular, $\phi_i^z(t)$ = 1, 2, and 3 indicates binary phase-shift keying (BPSK), quadrature-phase shift keying (QPSK), and 8 phase-shift keying (8PSK), respectively, and $\phi_i^z(t) \geq 4$ provides $2^{\phi_i^z(t)}$ quadrature amplitude modulation (QAM). 

Suppose that the BER requirement is $\varepsilon$. Consider the generic Nakagami-$m$ fading channel model, $\varepsilon$ is given by~\cite{alouini2000adaptive}
\begin{align}
&\varepsilon \approx \frac{0.2}{\Gamma(m)} (\frac{m}{\bar{\gamma}_i})^m \Big[\frac{\Gamma(m,b_{\phi_i^z(t)}\gamma_i(\phi_i^z(t)))}{(b_{\phi_i^z(t)})^m} - \nonumber \\
&~~~~~~~~~~~~~~~~~~~~~~~\frac{\Gamma(m,b_{\phi_i^z(t)}\gamma_i(\phi_i^z(t)+1))}{(b_{\phi_i^z(t)})^m}\Big], \\
&b_{\phi_i^z(t)} = \frac{m}{\bar{\gamma}_i} + \frac{3}{2(2^{\phi_i^z(t)}-1)},
\end{align}
where $\Gamma(\cdot)$ is the Gamma function~\cite{gradshteyn2014table}, and $\bar{\gamma}_i$ is the average SNR. $\gamma_i(\phi_i^z(t))$ is the SNR between device $i$ and the UAV using $\phi_i^z(t)$, as given by 
\begin{align}
\gamma_i(\phi_i^z(t)) = \frac{\|\mathbf{h}_i^z(t)\|^2 P^z_i(t)}{\sigma_0^2},
\end{align}
where $P^z_i(t)$ is the transmit power of device $i$, and $\sigma_0^2$ is noise power at the UAV. 
In particular, for illustration convenience, we consider a special case of the Nakagami-$m$ model in this paper where $m = 1$~\cite{li2015epla,wang2017pele,li2016reliable}. Note that the proposed deep reinforcement learning approach is generic, and can work
with other Nakagami fading channel model with any $m$ values. 
The required transmit power of the ground device depends on $\phi_i^z(t)$ and $\mathbf{h}_i^z(t)$, and can be given by~\cite{li2016energy}
\begin{align}
P_i^z(t)\approx\frac{\kappa_2^{-1}\ln\frac{\kappa_1}{\varepsilon}}{\|\mathbf{h}_i^z(t)\|^2}(2^{\phi_i^z(t)}-1),
\label{eq_txPower}
\end{align}
where $\kappa_1$ and $\kappa_2$ are channel related constants. 

According to~\cite{li2015poster}, the MPT efficiency is jointly decided by the distance between the MPT transmitter and the receiver, and their antenna alignment. Therefore, the power transferred to device $i$ at $t$ in lap $z$ via MPT is given by
\begin{align}
\tilde{P}^z_i(t) = \omega(d_i^z(t),\theta_i^z(t)) P_{\rm UAV}^{tx} \|\mathbf{h}_i^z(t)\|^2,
\label{eq_powerTx}
\end{align}
where $\|\cdot\|$ stands for norm. $\omega(d_i^z(t),\theta_i^z(t))$ is the MPT efficiency factor given the distance $d_i^z(t)$ and the MPT transceiver alignment $\theta_i^z(t)$ between the ground device and the UAV. The transmit power of the UAV is fixed and set to be $P_{\rm UAV}^{tx}$, so that the operations at the UAV can keep simple. 

Each of the wirelessly powered ground devices harvests energy from the UAV. The rechargeable battery is finite with the capacity of $E$ Joules, and the battery overflows if overcharged. Moreover, the battery readings are continuous variables with variance difficult to be traced in real-time. Therefore, to improve the mathematical tractability of the problem and for illustration convenience, the continuous battery is discretized into $K$ levels, as $0 <\mathcal {E} <2\mathcal {E}<\cdots<K\mathcal {E}=E$. $e_{i,z}(t)\in\{0,\mathcal {E},2\mathcal {E},\cdots,K\mathcal {E}\}$~\cite{liu2014selection}. In other words, the battery level of the ground device is lower rounded to the closest discrete level. 

We also assume that all the ground devices have the same battery size, data queue size, packet length, and the BER requirement. However, our proposed resource allocation approach can be extended to a heterogeneous network setting, where the complexity of the resource allocation problem may grow as the result of an increased number of MDP states. 

\subsection{Communication protocol}
\label{sec_protocol}
\begin{figure}[htb]
\centering
\includegraphics[width=3.5in]{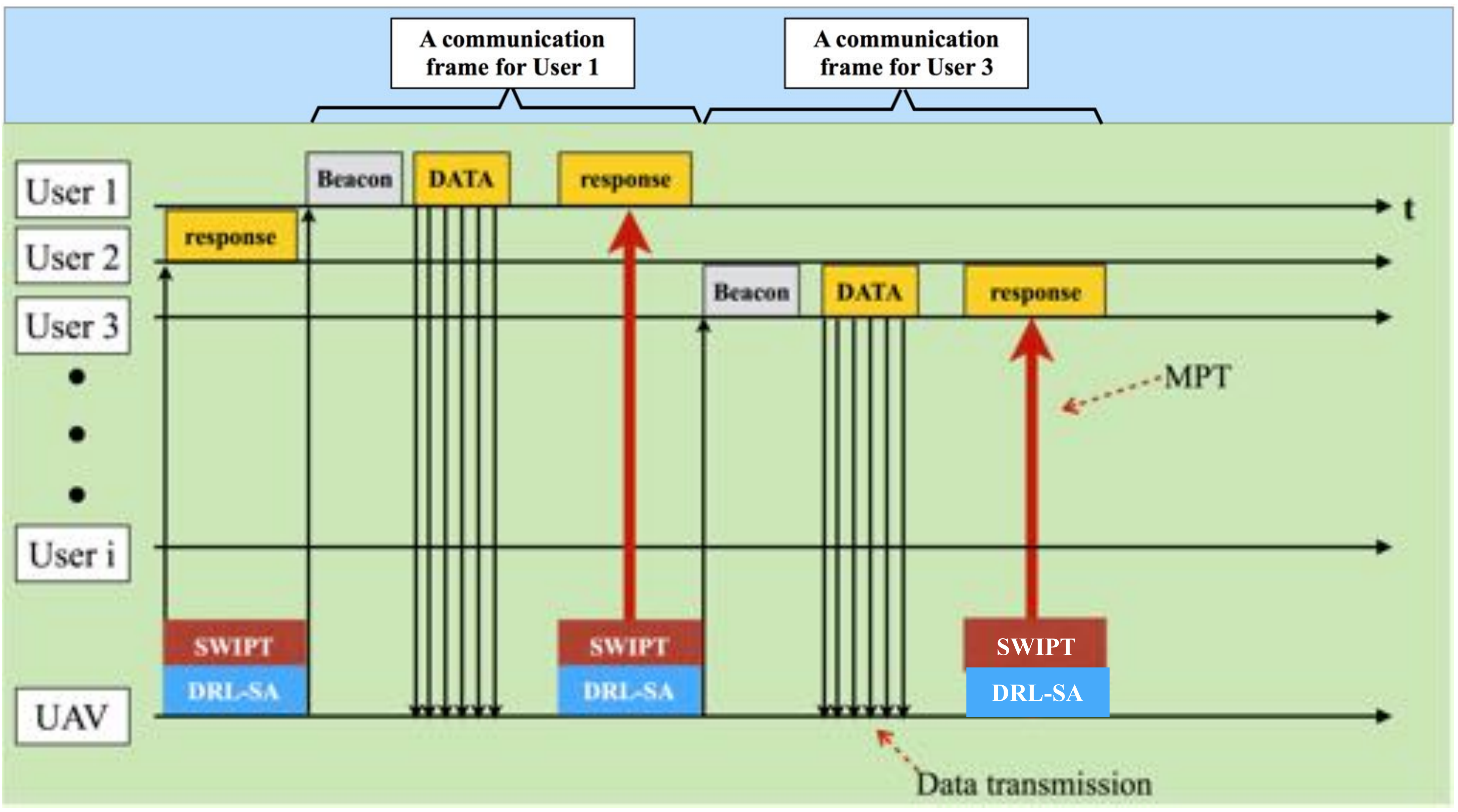}
\caption{\small{The communication protocol for UAV-assisted online MPT and data collection.} }
\label{fig_protocol}
\end{figure}

Figure~\ref{fig_protocol} illustrates the communication protocol for UAV-assisted online MPT and data collection. Specifically, each communication frame that contains a number of time slots is allocated to the ground device for MPT and data transmission. The ground device selection is determined by the UAV, using DRL-SA which will be illustrated in Section~\ref{sec_DRL}. Then, the UAV broadcasts a short beacon message to the selected ground device for data transmission. On the reception of the ground device's data, the UAV is aware of the information for MPT, i.e., $d_i^z(t)$, $\theta_i^z(t)$, and $\mathbf{h}_i^z(t)$. Moreover, a control segment of the device's data packet contains $q_{i,z}(t)$ and $e_{i,z}(t)$. The overhead of this control segment is small. For example, consider $e_{i,z}(t)$ of 100 and $q_{i,z}(t)$ of 100 packets, the overhead is only 12 bits, much smaller than the size of the data packet. Therefore, we assume that the transmission time and the energy consumption of the control segment are negligible. 

The UAV processes the received data packets online, and responds to the ground device's requests, e.g., providing network access services, or online information query. SWIPT is utilized to transmit the response to the ground device and charge its battery via MPT simultaneously. Meanwhile, DRL-SA is conducted by the UAV to schedule the other ground device for MPT and data transmission in the next communication frame.

%=============================================================================%
%============================Section 4 Deep Reinforcement Learning===================%
\section{Deep Reinforcement Learning for UAV-assisted online MPT and Data Collection}
\label{sec_DRL}
In this section, we first present the problem formulation based on MDP, and provide a reinforcement learning solution to the problem. 
Due to the curse-of-dimensionality of reinforcement learning, we propose a new deep reinforcement learning based scheduling algorithm to minimize overall data packet loss of the ground devices, by optimally deciding the device to be charged and interrogated for data collection, and the instantaneous patrolling velocity of the UAV. 
The modulation scheme of the ground device is also optimized to maximize the harvested energy. 
Finally, the optimality of the deep reinforcement learning approach is analyzed. 

\subsection{MDP formulation}
The UAV takes the actions of the ground device selection, the modulation scheme, and the instantaneous patrolling velocity of the UAV. Each action depends on the current network state, i.e, the battery level $e_{i,z}(t)$ and queue length $q_{i,z}(t)$ of every device $i$, the channel quality $\mathbf{h}_i^z(t)$, and the location of the UAV $\zeta_z(t)$ along the flight trajectory. The actions also account for the potential influence on the future evolution of the network. Particularly, the current action that the UAV takes can affect the future battery level and queue length of every device and, in turn, influence the future actions to be taken. Such actions are a discrete-time stochastic control process which is partly random (due to the random and independent arrival/queueing process of sensory data at every device) and partly under the control of the decision-making UAV. 

The action can be optimized in a sense that the optimality in regards of a specific metric, e.g., packet loss from queue overflows and unsuccessful data transmissions, is achieved in the long term over the entire stochastic control process (rather than myopically at an individual time slot). Motivated by this, we consider an MDP formulation for which the actions are chosen in each state to minimize a long-term objective. 
An MDP is defined by the quadruplet $\langle$$\mathcal{S}$, $\mathcal{A}$, $C\Big\{\mathcal{S}_\beta\Big|\mathcal{S}_\alpha,k\Big\}$, $\Pr\Big\{\mathcal{S}_\beta\Big|\mathcal{S}_\alpha,k\Big\}$$\rangle$, where $\mathcal{S}$ is the set of possible states; $\mathcal{A}$ is the set of actions; $C\Big\{\mathcal{S}_\beta\Big|\mathcal{S}_\alpha,k\Big\}$ is the immediate cost yielded when action $k$ is taken at state $\mathcal{S}_\alpha$ and the following state changes to $\mathcal{S}_\beta$; and $\Pr\Big\{\mathcal{S}_\beta\Big|\mathcal{S}_\alpha,k\Big\}$ denotes the transition probability from state $\mathcal{S}_\alpha$ to state $\mathcal{S}_\beta$ when action $k$ is taken.

The resource allocation problem of interest in UAV-assisted online MPT and data collection can be formulated as a discrete-time MDP, where each state $\mathcal{S}_\alpha$ collects the battery levels and queue lengths of the ground devices, the channel quality between the UAV and device $i$, and the location of the UAV, i.e., $\{(q_{i,z}(t), e_{i,z}(t), \mathbf{h}_i^z(t), \zeta_z(t)), i = 1,2,...,I; z = 1,2,...,Z\}$. The size of the state space, i.e., the number of such states, is $\Big((K+1)(D+1)\Big){I Z} + HZ + VZ$, where $H$ is the number of channel states, and $V$ is the number of waypoints on the UAV's trajectory. The action $\mathcal{A}$ to be taken is to schedule one device to transmit data to the UAV at time slot $t$, while specifying the modulation of the device and the instantaneous patrolling velocity of the UAV, i.e., $\mathcal{A} \in \Big\{ (i,\phi_i^z(t), v^z(t)):  i = 1,2,...,I; z = 1,2,...,Z; \phi_i^z(t) \in \{1,2,...,\Phi\};v_{\rm min} \leq v^z(t) \leq v_{\rm max} \Big\}$. The size of the action set is $I Z | v^z(t) |$, where $| v^z(t) |$ stands for the cardinality of the set $[v_{\rm min},v_{\rm max}]$. 

\begin{figure}[htb]
\centering
\includegraphics[width=3.5in]{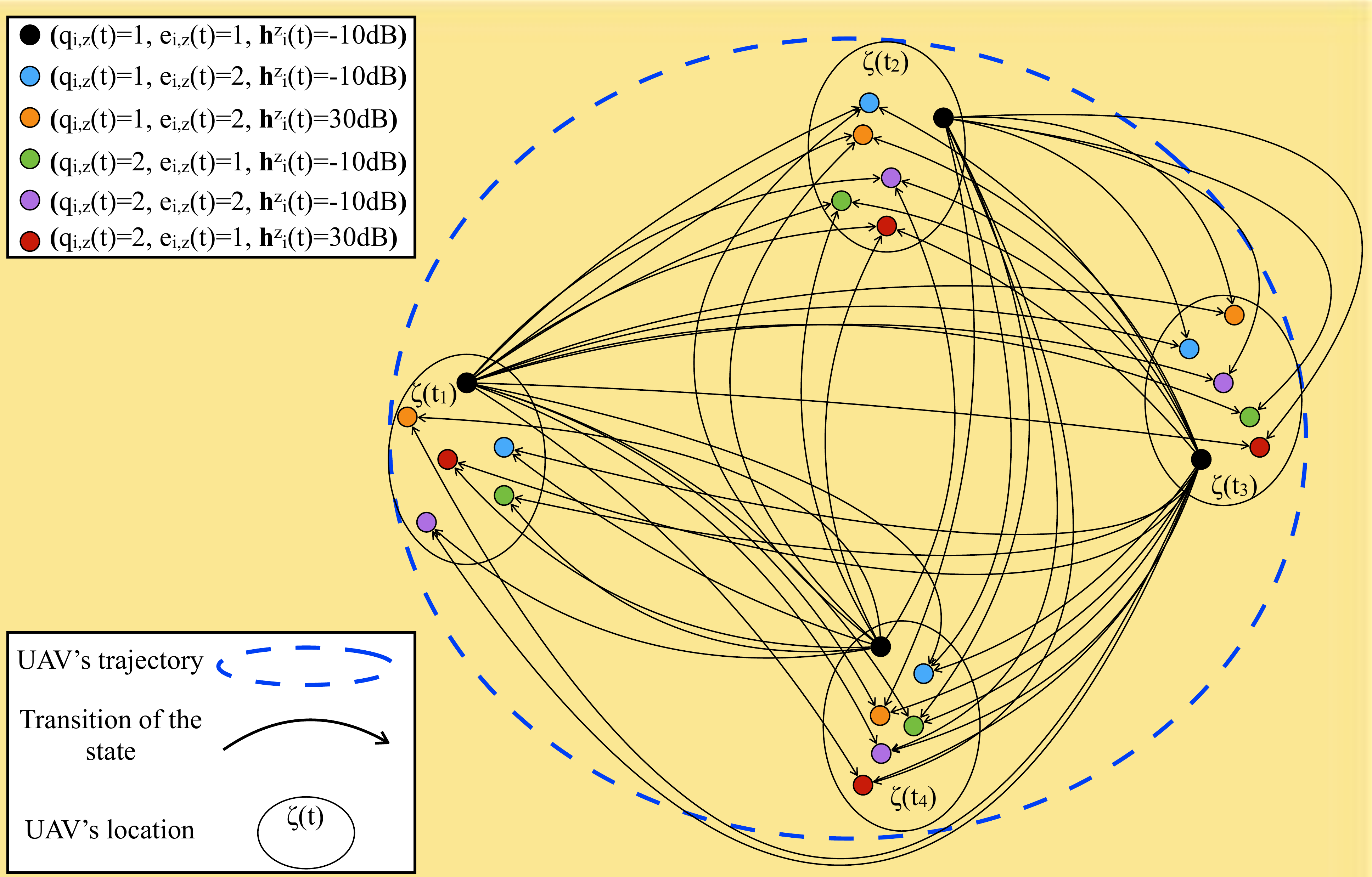}
\caption{\small{An example of a transition diagram of 24 MDP states, which has 4 waypoints on the UAV's trajectory.} }
\label{fig_mdp}
\end{figure}

To illustrate the proposed MDP model, Figure~\ref{fig_mdp} presents an example of transition diagram with 24 MDP states in one lap of the UAV's flight, where $I = 1$, $Z = 1$, $K = 1$, $D = 1$, $H = 2$ (e.g., $-10$dB, $30$dB), and $V = 4$. 
The vertices stand for all possible states in MDP, i.e., $\{(q_{i,z}(t), e_{i,z}(t), \mathbf{h}_i^z(t), \zeta_z(t))\}$. The edges show the transition from each state to other states according to $\Pr\Big\{\mathcal{S}_\beta\Big|\mathcal{S}_\alpha,k\Big\}$. 
The state transition depends on the change of $\{(q_{i,z}(t), e_{i,z}(t), \mathbf{h}_i^z(t)\}$ of the ground device and $\zeta_z(t)$ along the trajectory of the UAV. In other words, the next state of $\{(q_{i,z}(t_1), e_{i,z}(t_1), \mathbf{h}_i^z(t_1), \zeta_z(t_1))\}$ can be one of the states at $\zeta_z(t_2)$, $\zeta_z(t_3)$, or $\zeta_z(t_4)$. 
For example, for $t_1$, the next state of $\{(q_{i,z}(t_1) = 1, e_{i,z}(t_1) = 1, \mathbf{h}_{i}^{z}(t_1) = -10{\rm dB}, \zeta_{z}(t_1))\}$ can be $\{(q_{i,z}(t_2) = 2, e_{i,z}(t_2) = 2, \mathbf{h}_{i}^{z}(t_2) = -10{\rm dB}, \zeta_{z}(t_2))\}$, if device $i$ is selected, but the data collection is not successful; or $\{(q_{i,z}(t_2) = 1, e_{i,z}(t_2) = 2, \mathbf{h}_{i}^{z}(t_2) = -10{\rm dB}, \zeta_{z}(t_2))\}$, if the data collection is successful. 
Note that Figure~\ref{fig_mdp} gives a small-scale example of the transition of one of the states, i.e., $\{(q_{i,z}(t) = 1, e_{i,z}(t) = 1, \mathbf{h}_{i}^{z}(t) = -10{\rm dB}, \zeta_{z}(t))\}$. 
The UAV's trajectory can have hundreds of waypoints and the model can contain over $8 \times 10^5$ MDP states, as configured in Section~\ref{sec_evaluation}, which leads to an extremely complex state transition diagram. 

The optimal policy in MDP can be determined by classical approaches, e.g., value iteration (computes and improves the action-value function estimate iteratively) or policy iteration (redefines the policy at each step and computes the value according to this new policy). However, these two methods require that the transition probability and the cost of all states have been accurately known. In contrast, this paper is interested in a practical scenario where the UAV has no a-prior knowledge on $C\Big\{\mathcal{S}_\beta\Big|\mathcal{S}_\alpha,k\Big\}$ and $\Pr\Big\{\mathcal{S}_\beta\Big|\mathcal{S}_\alpha,k\Big\}$. 

\subsection{Q-learning}
Q-learning, one of the reinforcement learning techniques, can obtain the optimal resource allocation when the transition and/or cost functions are unknown while minimizing the long-term expected accumulated discounted costs (i.e., the expected packet loss of the ground devices)~\cite{li2018reinforcement}. 
We define the objective function of the resource allocation problem as $v(S_\alpha)$, which gives 
\begin{equation}
v(S_\alpha) = \min_{\pi \in \Pi}~\mathbb{E}^\pi_{\mathcal{S}_\alpha} \left\{ \sum^{\infty}_{t=1} \delta^{t-1} C\Big\{\mathcal{S}_\beta\Big|\mathcal{S}_\alpha,k\Big\}\right\}, 
\label{eq_max_1}
\end{equation}
where $\delta \in [0, 1]$ is a discount factor for future states. $C\Big\{\mathcal{S}_\beta \Big| \mathcal{S}_\alpha,k\Big\}$ is the cost from state $\mathcal{S}_\alpha$ to $\mathcal{S}_\beta$ when action $k$ is carried out. $\mathbb{E}^\pi_{\mathcal{S}_\alpha} \{\cdot\}$ denotes the expectation with respect to policy $\pi$ and state $\mathcal{S}_\alpha$. 
Furthermore, the action-value function $Q \Big\{ \mathcal{S}_\beta \Big| \mathcal{S}_\alpha, k \Big\}$ defines the expected cost after observing state $\mathcal{S}_\alpha$ and taking action $k$~\cite{mnih2015human}. 
By performing the optimal action $k^*$, the optimal action-value function can be expressed as a combination of the expected cost and the minimum value of $Q \Big\{ \mathcal{S}_{\beta^\prime} \Big| \mathcal{S}_\beta, k^\prime \Big\}$, where $\mathcal{S}_{\beta^\prime}$ is the next state of $\mathcal{S}_\beta$, and $k^\prime$ is the next action of $k$. Thus, we have  
\begin{equation}
\begin{array}{c l}
&Q^* \Big\{\mathcal{S}_\beta \Big| \mathcal{S}_\alpha,k^* \Big\} = (1-\varrho) Q^* \Big\{\mathcal{S}_\beta \Big| \mathcal{S}_\alpha,k^*\Big\} +\\
&~~\varrho \Big[ C\Big\{\mathcal{S}_\beta\Big|\mathcal{S}_\alpha,k^* \Big\} + \delta \min_{k^\prime \in \mathcal{A}} Q \Big\{ \mathcal{S}_{\beta^\prime} \Big| \mathcal{S}_\beta, k^\prime \Big\} \Big]. 
\end{array}
\label{eq_optQfunc}
\end{equation} 
where $\varrho \in (0,1]$ (a small positive fraction) indicates learning rate. As observed in~\eqref{eq_optQfunc}, the convergence rate of $Q \Big\{\mathcal{S}_\beta \Big| \mathcal{S}_\alpha,k\Big\}$ to $Q^* \Big\{\mathcal{S}_\beta \Big| \mathcal{S}_\alpha,k^* \Big\}$ depends on the discount factor $\delta$. Namely, the convergence rate of the action-value function increases with the discount factor $\delta$. 

However, Q-learning suffers from the well-known curse of dimensionality. Thus, Q-learning is impractical for the resource allocation problem in the UAV-assisted online MPT and data collection due to a large state and action space in which the time needed for Q-learning to converge grows immeasurably. In addition, Q-learning is unstable when combined with non-linear approximation functions such as neural networks~\cite{tsitsiklis1997analysis}. 

\subsection{Deep reinforcement learning based approach}
To circumvent the curse-of-dimensionality of reinforcement learning, we propose an on-board deep Q-network for the UAV to optimize the resource allocation online by approximating the optimal action-value function. 
\begin{figure}[htb]
\centering
\includegraphics[width=3.5in]{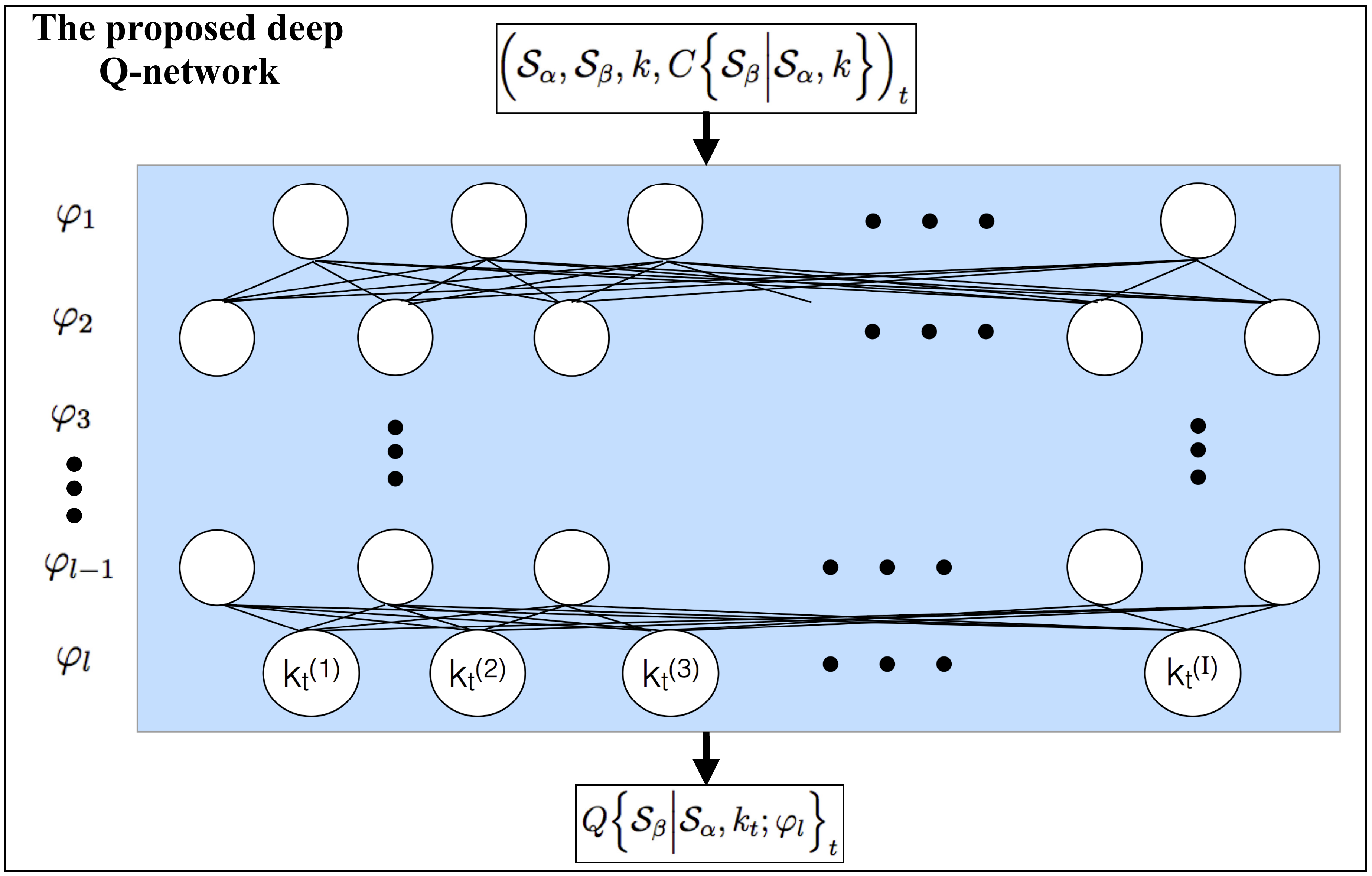}
\caption{\small{Schematic illustration of the proposed on-board deep Q-network.} }
\label{fig_deepQnetwork}
\end{figure}
As shown in Figure~\ref{fig_deepQnetwork}, the action-value function in~\eqref{eq_optQfunc} is represented by the proposed on-board deep Q-network, which takes the current network state and action as the input and derives the corresponding action-value function. 
Given the total number of iterations $\Omega$, $Q \Big\{\mathcal{S}_\beta \Big| \mathcal{S}_\alpha,k;\varphi_l\Big\}$ is approximated by adapting a set of weights $\varphi_l$, where $l \leq \Omega$. 
By optimizing $\varphi_l$, the outputs of the deep Q-network, i.e., the approximated $Q \Big\{\mathcal{S}_\beta \Big| \mathcal{S}_\alpha,k;\varphi_l\Big\}$, can be minimized. 

\begin{algorithm}[t]
\caption{Deep Reinforcement Learning based Scheduling Algorithm}
\label{alg_drlsa}
\begin{algorithmic}[1]
\STATE{\textbf{1. Initialize}: }
\STATE{$\mathcal{S}_\alpha \in {\cal S}$, $k \in \mathcal{A}$, $\varrho$, and $w$ $\to Q \Big\{\mathcal{S}_\beta \Big| \mathcal{S}_\alpha,k\Big\}$.}
\STATE{Random weights $\varphi_l \to Q \Big\{\mathcal{S}_\beta \Big| \mathcal{S}_\alpha,k; \varphi_l\Big\}$.}
\STATE{Weights $\varphi_{l-1} = \varphi_l$ $\to$ target deep Q-network $\hat{Q} \Big\{\mathcal{S}_\beta \Big| \mathcal{S}_\alpha,k; \varphi_{l-1}\Big\}$.}
\STATE{Learning time $\to t_{\rm learning}$.}
\STATE{\textbf{2. Learning}: }
\FOR{$episode = 1 \to M$}
\STATE{Start state $\mathcal{S}_\alpha \to \mathcal{S}_1$ and accordingly update $\varphi_l \to \varphi_1$.}
\FOR{$t = 1 \to t_{\rm learning}$}
\IF{Probability $\epsilon$}
\STATE{Select a random action $k_t$.}
\ELSE
\STATE{$k_t \to \text{argmin}_k Q \Big\{ \mathcal{S}_\beta \Big| \mathcal{S}_\alpha, k;\varphi_l \Big\}$}
\ENDIF
\STATE{Execute action $k_t$ in the environment.} 
\STATE{Obtain the cost $C\Big\{\mathcal{S}_\beta\Big|\mathcal{S}_\alpha,k_t\Big\}$ and the next state $\mathcal{S}_\beta$ at $t+1$.}
\STATE{Store transition $\Big(\mathcal{S}_\alpha, \mathcal{S}_\beta, k_t, C\Big\{\mathcal{S}_\beta\Big|\mathcal{S}_\alpha,k_t\Big\}\Big)_t \to \mathbb{D}[\cdot]$.}
\STATE{Sample random minibatch $\Big(\mathcal{S}_\alpha, \mathcal{S}_\beta, k, C\Big\{\mathcal{S}_\beta\Big|\mathcal{S}_\alpha,k\Big\}\Big)_l$ from $\mathbb{D}[\cdot]$.}
\IF{$\mathcal{S}_\beta$ terminates at step $l+1$}
\STATE{Set $y_l \to C\Big\{\mathcal{S}_\beta\Big|\mathcal{S}_\alpha,k_t\Big\}_l$.}
\ELSE
\STATE{Set $y_l \to C\Big\{\mathcal{S}_\beta\Big|\mathcal{S}_\alpha,k_t\Big\}_l +  \delta \min_{k^\prime \in \mathcal{A}} \hat{Q} \Big\{ \mathcal{S}_{\beta^\prime} \Big| \mathcal{S}_\beta, k^\prime;\varphi_{l-1} \Big\}$}.
\ENDIF
\STATE{Derive the loss function $\Gamma_l(\varphi_l) \to (y_l - Q \Big\{\mathcal{S}_\beta \Big| \mathcal{S}_\alpha,k_t;\varphi_l \Big\}_t)^2$.}
\STATE{Perform~\eqref{eq_gradient} $\to$ gradients with respect to $\varphi_l$.}
\IF{There are $U$ number of learning updates.}
\STATE{Synchronize $\varphi_{l-1} \to \varphi_l$.} 
\STATE{Reset $\hat{Q} \Big\{\mathcal{S}_\beta \Big| \mathcal{S}_\alpha,k; \varphi_{l-1}\Big\} \to Q \Big\{\mathcal{S}_\beta \Big| \mathcal{S}_\alpha,k_t; \varphi_l\Big\}_t$.}
\ENDIF
\ENDFOR
\ENDFOR
\STATE{\textbf{3. Modulation}: }
\STATE{Determining $\phi^z_k(t)^{\star}$ for the scheduled ground device $k$ (see Appendix).}
\end{algorithmic}
\end{algorithm}

In our on-board deep Q-network, experience replay is conducted to randomize over the states and the actions of MDP at each time-step $t$, i.e., $\Big(\mathcal{S}_\alpha, \mathcal{S}_\beta, k, C\Big\{\mathcal{S}_\beta\Big|\mathcal{S}_\alpha,k\Big\}\Big)_t$, thereby removing correlations in the observation sequence and smoothing over changes in the data distribution. Specifically, $\Big(\mathcal{S}_\alpha, \mathcal{S}_\beta, k, C\Big\{\mathcal{S}_\beta\Big|\mathcal{S}_\alpha,k\Big\}\Big)_t$ is stored in a data set $\mathbb{D}[\cdot]$, pooled over many episodes (where an episode ends when a terminal state is reached) into an experience replay memory. 
Moreover, samples (or minibatches) of the experience in the deep Q-network are accordingly updated during learning. 
The deep Q-network is trained by adjusting the weight $\varphi_l$ at iteration $l$ so as to minimize a sequence of loss functions $\Gamma_l(\varphi_l)$; see~\eqref{eq_lossfunction}, 
\begin{figure*}
\begin{align}
\Gamma_l(\varphi_l) = \mathbb{E}_{\Big(\mathcal{S}_\alpha,\mathcal{S}_\beta,k,C\Big\{\mathcal{S}_\beta\Big|\mathcal{S}_\alpha,k\Big\}\Big) \sim \mathbb{D}\Big[\Big(C\Big\{\mathcal{S}_\beta\Big|\mathcal{S}_\alpha,k\Big\} + \delta \min_{k^\prime \in \mathcal{A}} Q \Big\{ \mathcal{S}_{\beta^\prime} \Big| \mathcal{S}_\beta, k^\prime;\varphi_{l-1} \Big\} - Q \Big\{ \mathcal{S}_\beta \Big| \mathcal{S}_\alpha, k;\varphi_l \Big\} \Big)^2\Big]}, 
\label{eq_lossfunction}
\end{align}
\end{figure*}
where $\varphi_l$ and $\varphi_{l-1}$ are the weights at iterations $l$ and $(l-1)$, respectively. 
At each iteration of minimizing $\Gamma_l(\varphi_l)$, the weight $\varphi_{l-1}$ from iteration $(l-1)$ is fixed. Thus, the subproblem of learning $\Gamma_l(\varphi_l)$ at iteration $l$ $(l \leq \Omega)$ defines $C\Big\{\mathcal{S}_\beta\Big|\mathcal{S}_\alpha,k\Big\} + \delta \min_{k^\prime \in \mathcal{A}} Q \Big\{ \mathcal{S}_{\beta^\prime} \Big| \mathcal{S}_\beta, k^\prime;\varphi_{l-1} \Big\}$. 
For each sample (or minibatch), the current weight $\varphi_l$, gradient descent is learned to derive weights $\varphi_l$, which iteratively computes the gradient $\triangledown\Gamma_l(\varphi_l)$, and updates the neural network's weights to reach the global minimum. We differentiate $\Gamma_l(\varphi_l)$ with regards to $\varphi_l$, and obtain~\eqref{eq_gradient}. 
\begin{figure*}
\begin{align}
\triangledown\Gamma_l(\varphi_l) = \mathbb{E}_{\Big(\mathcal{S}_\alpha,\mathcal{S}_\beta,k,C\Big\{\mathcal{S}_\beta\Big|\mathcal{S}_\alpha,k\Big\}\Big)} \Big[\Big(&C\Big\{\mathcal{S}_\beta\Big|\mathcal{S}_\alpha,k\Big\} + \delta \min_{k^\prime \in \mathcal{A}} Q \Big\{ \mathcal{S}_{\beta^\prime} \Big| \mathcal{S}_\beta, k^\prime;\varphi_{l-1} \Big\} - \nonumber \\
&Q \Big\{ \mathcal{S}_\beta \Big| \mathcal{S}_\alpha, k;\varphi_l \Big\} \Big) \triangledown_{\varphi_l} Q \Big\{ \mathcal{S}_\beta \Big| \mathcal{S}_\alpha, k;\varphi_l \Big\} \Big]. 
\label{eq_gradient}
\end{align}
\hrulefill
\end{figure*}

Algorithm~\ref{alg_drlsa} presents the proposed DRL-SA scheme, which optimizes the actions based on the deep Q-network to solve the online resource allocation problem. Specifically, $U$ is a global parameter for counting the total number of updates to the UAV, rather than counting the updates from the local learner. The $\epsilon$-greedy policy is utilized to balance the action-value function minimization based on the knowledge already known, as well as trying new actions to obtain knowledge unknown. 
In particular, the optimal action of determining $\phi_i^z(t)$ for maximizing the harvested energy (see the Appendix) is carried out once the optimal ground device is selected from the deep Q-network. 

Note that the optimal solutions to a small-scale MDP problem of interest can be solved by Q-learning in which the series of the optimal actions are obtained in a state machine, given the optimal decisions at each of the states. However, Q-learning is impractical for the complex resource allocation problem that contains a large state and action space, due to the well-known curse of dimensionality. 
As observed in Algorithm~\ref{alg_drlsa}, our on-board deep Q-network maintains two separate Q-networks $Q \Big\{ \mathcal{S}_\beta \Big| \mathcal{S}_\alpha, k;\varphi_l \Big\}$ and $Q \Big\{ \mathcal{S}_\beta \Big| \mathcal{S}_\alpha, k;\varphi_{l-1} \Big\}$ with current weights $\varphi_l$ and the old weights $\varphi_{l-1}$, respectively~\cite{van2012reinforcement}. $\varphi_l$ can be updated many times per time-step, and copied into $\varphi_{l-1}$ after every $U$ iterations. At every update iteration, the deep Q-network is trained to minimize the mean-squared Bellman error, by minimizing the loss function $\Gamma_l(\varphi_l)$. 
Therefore, the proposed DRL-SA can achieve the optimality asymptotically, with the growing size of the deep Q-network.

%=============================================================================%
%============================Section 5 Perfect Markov Decision=======================%
%\input{secDPA}

%=============================================================================%
%============================Section 6 Evaluation=================================%
\section{Numerical Results and Discussions}
\label{sec_evaluation}
In this section, we first present network configurations and performance metrics. Then, we evaluate the network cost of the proposed DRL-SA scheme with regards to the network size, data queue length, and learning discount factors. Here, the network cost defines the amount of packet loss due to the data queue overflow and the failed transmission from the ground device to the UAV. 
We also show the UAV's velocity in terms of the network size, as well as the data queue length. 

\subsection{Implementation of DRL-SA}
\label{sec_config}
The simulation platform is an INSYS Server running 64-bit Ubuntu 16.04 LTS, with 4-core Intel i7-6700K 4GHz CPUs and 16G memory. 
DRL-SA is implemented in Python 3.5 by using Google TensorFlow~\cite{abadi2016tensorflow} (the symbolic math library for numerical computation) with Keras~\cite{chollet2015keras} (the Python deep learning library). 
We develop the on-board deep Q-network in DRL-SA according to the following steps: 
\begin{itemize}
\item Initialize the configurations. Each device generates 100 data packets, where the packet length is 128 bytes. The transmit power of the UAV is 100 milliwatts. $\varepsilon$ is set to 0.05\%, however, the value of $\varepsilon$ can be configured depending on the traffic type and quality-of-service (QoS) requirement of the user's data, as well as the transmission capability of the UAV. Other simulation parameters are listed in Table~\ref{tb_config}. 
\begin{table}[htb]
    \centering
    \caption{TensorFlow configurations}
    \begin{tabular} {|p{5cm}|p{4cm}|} \hline
        \bf{Parameters} & \bf{Values} \\ \hline
        Number of ground devices & 50 $\backsim$ 200 \\ \hline
        Queue length & 20 $\backsim$ 60 \\ \hline
        Energy levels & 50 \\ \hline
        Number of UAV's waypoints & 100 \\ \hline
        Discount factor &  0.99  \\ \hline
	Learning rate &  0.0001  \\ \hline 
	Replay memory size & 5000 \\ \hline
	Batch size & 32  \\ \hline 
	Number of steps &  1000 \\ \hline 
	Number of episodes & 500 \\ \hline
   \end{tabular}
\label{tb_config}
\end{table}
\item Set up the architecture of the deep Q-network. Three fully connected hidden layers are created by using \textit{tensorflow.layers.dense(inputs, dimensionality of the output space, activation function)}. Then, \textit{tensorflow.train.AdamOptimizer().minimize(loss function)} is called to minimize the loss function. In particular, the optimizer is imported from the Keras library. 
\item Build the memory for the experience replay. For online training the deep Q-network, the memory stores the learning outcomes, \textit{a.k.a} experience at every step, using the quadruplet \textit{$<$state, action, cost, next\_state$>$}. The deep Q-network updates the memory by calling the function \textit{memory.add\_sample((state, action, cost, next\_state))}, and retrieves the experiences by using \textit{memory.sample(batch size)}. 
\item Create a deep Q-network agent, and kick off the learning. The agent is configured and compiled to take actions, and observe the cost and the new state. A TensorFlow session is created by implementing \textit{tensorflow.Session()} to execute the learning and evaluate the learning progress. 
\end{itemize}

For performance comparison, the proposed DRL-SA scheme is compared with two other on-board online scheduling policies as 
\begin{itemize}
\item Random scheduling policy (RSA). The UAV randomly schedules one ground device to collect data and transfer power at each time slot. The resource allocation is independent of battery and data queue of the ground device, channel variation, or UAV's trajectory. 
\item Longest queue scheduling policy (LQSA). This is a greedy algorithm, where the scheduling is based on the data queue length of the ground device. The device with the longest queue is given the highest priority to transmit data and harvest power. For LQSA, it is assumed that the up-to-date information of the data queue length at each device is known by the UAV. 
\end{itemize}
Both RSA and LQSA are implemented in Matlab. Their scheduling policies are not obtained by the deep Q-network. 
Note that the existing learning-based approaches, e.g., QSA~\cite{li2018reinforcement} or SARSA~\cite{hoang2017optimal}, can only solve the small-scale scheduling problem, and do not apply to the resource allocation in UAV-assisted online MPT and data collection. 
In addition, the resource allocation optimization schemes, e.g., EHMDP~\cite{li2018wireless} or EACH~\cite{zhang2013distributed}, require the a-prior knowledge about the network and have to be conducted in offline. 

\subsection{Performance evaluation}

\subsubsection{Network size}
\begin{figure}[htb]
\centering
\includegraphics[width=4in]{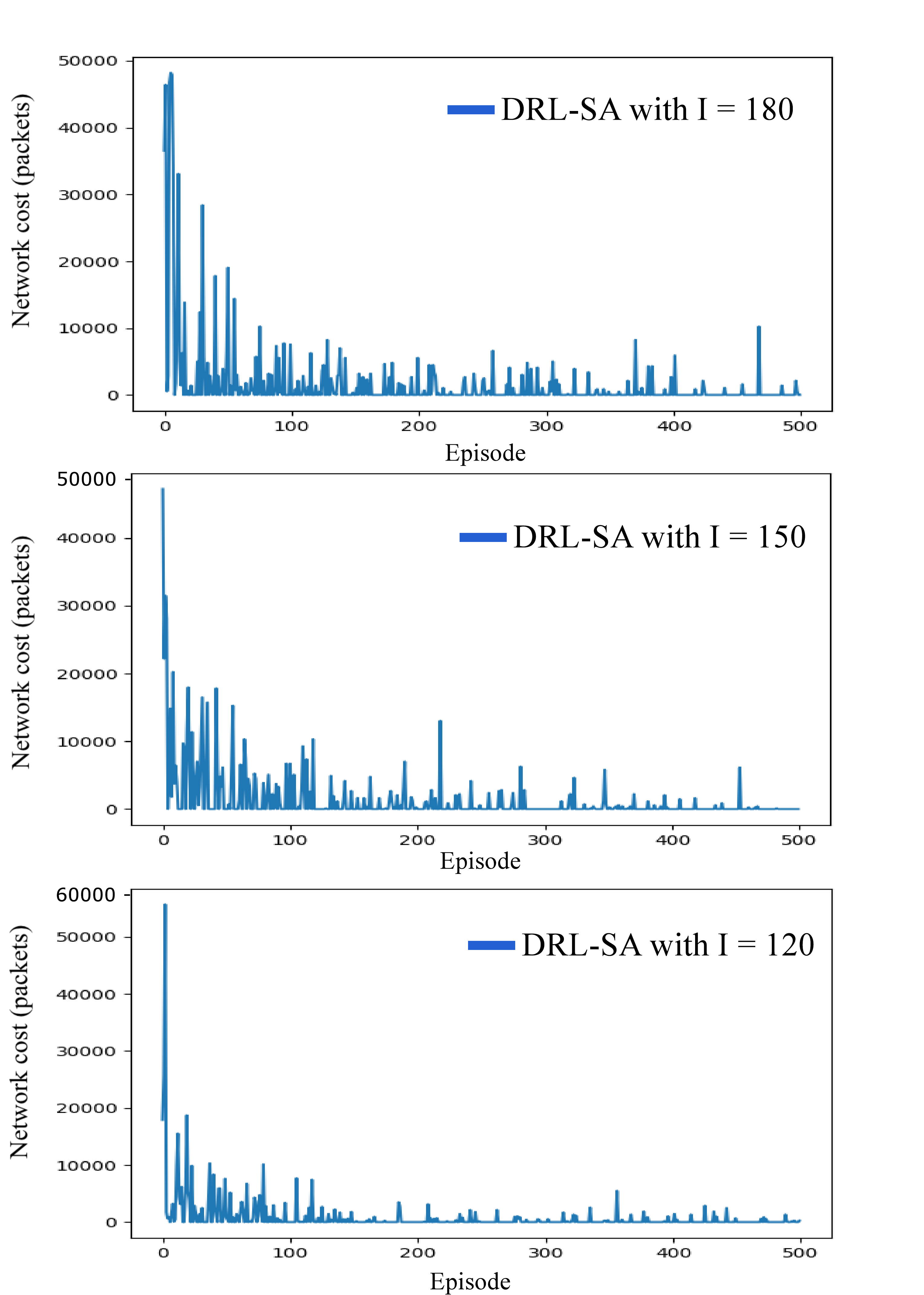}
\caption{\small{Network cost at the episode in terms of the different number of ground devices, i.e., $I$.} }
\label{fig_cost_episode}
\end{figure}

We assess the performance of DRL-SA when the number of ground devices enlarges from 50 to 200. 
Figure~\ref{fig_cost_episode} shows the network cost at each episode, given $I = 120, 150$ and $180$. The network cost of DRL-SA is high at the beginning of the learning process. With an increasing number of episodes, the network cost drops significantly until it reaches a relatively stable value. It confirms the fact that deep Q learning gradually converges after a number of episodes. 
Moreover, the average network cost is smaller when the number of ground devices is 120, as compared to the cases where the number of ground devices is 150 and 180, since the increasing number of ground devices leads the data queues to increasingly overflow. 

\begin{figure*}[htb]
\begin{center}
\begin{tabular}{ccc}
\includegraphics[width=3.2in]{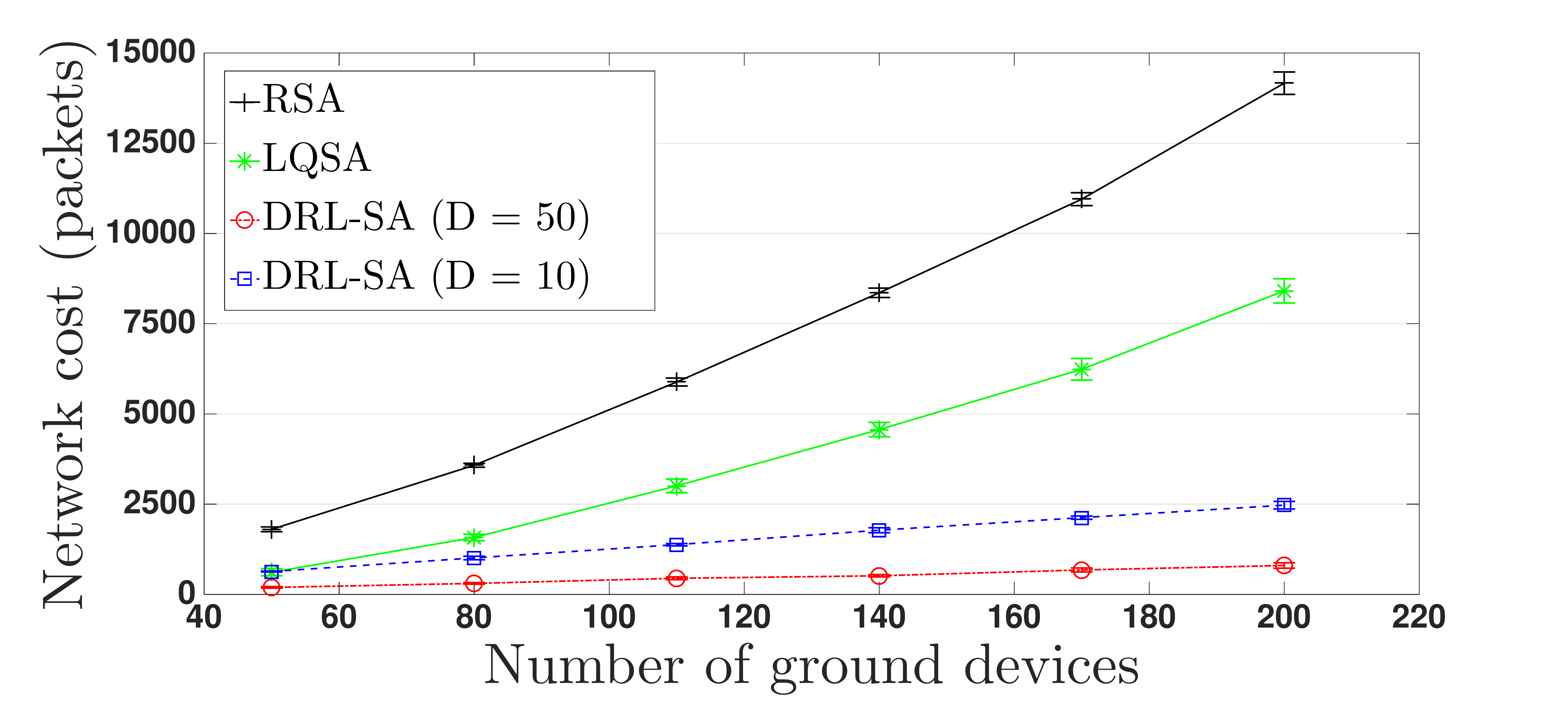} & \includegraphics[width=3.2in]{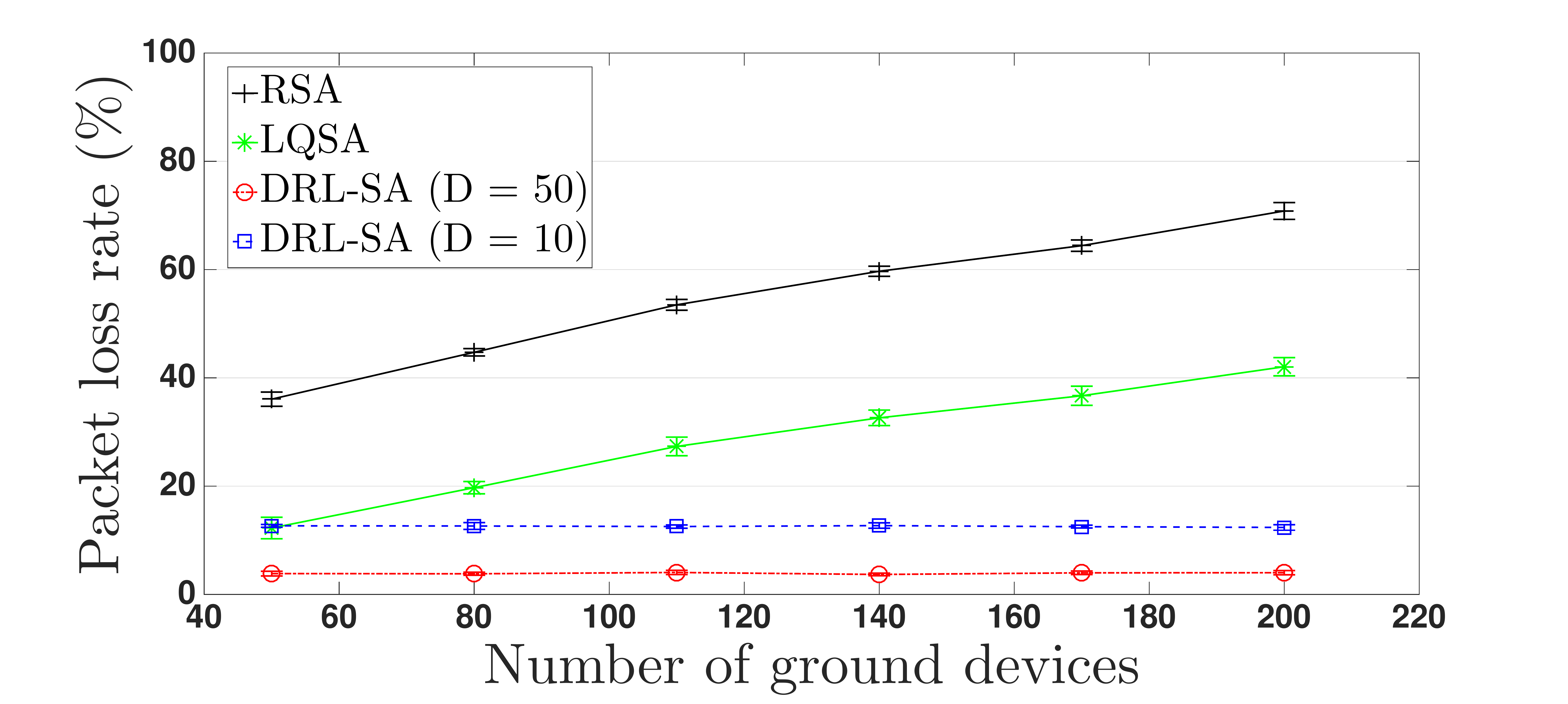} & \\\includegraphics[width=3.2in]{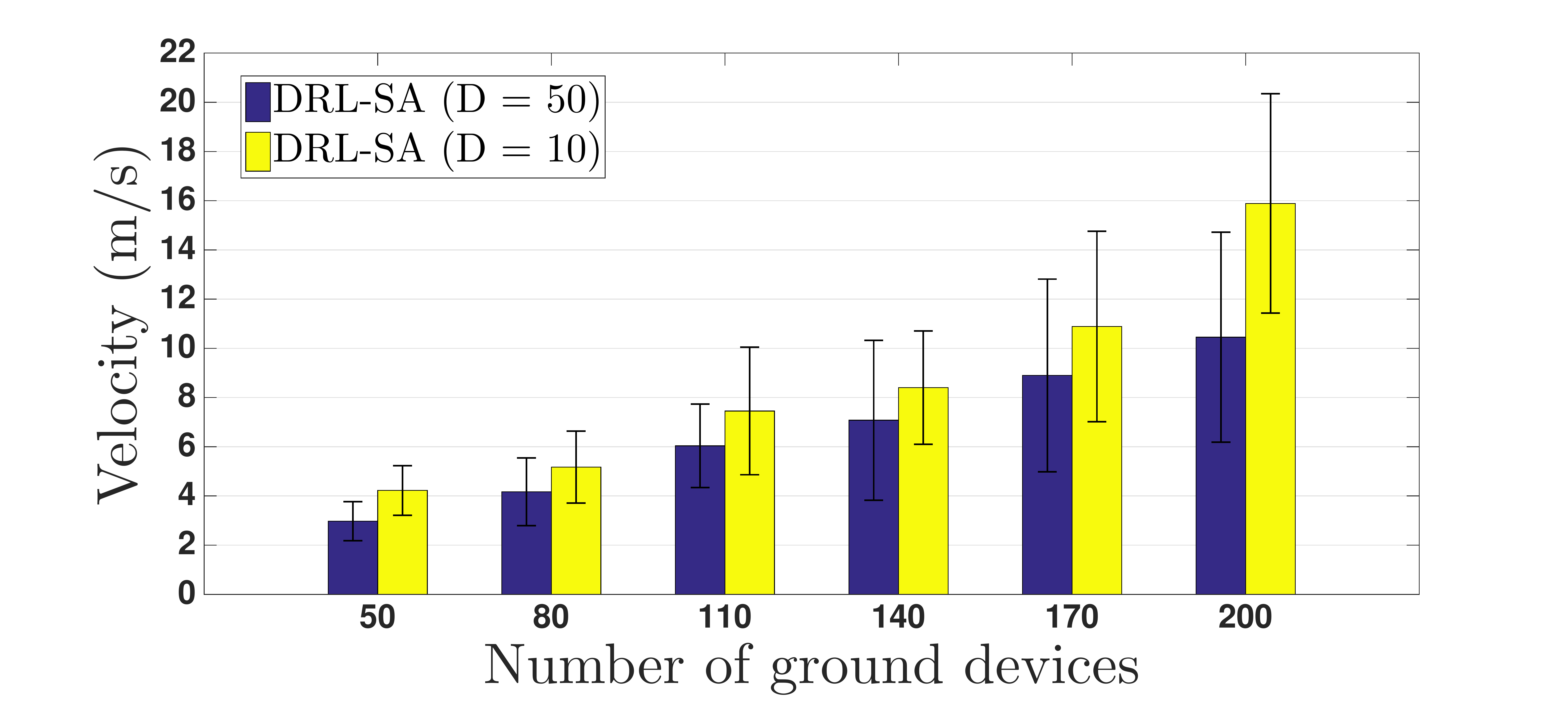} 
\\ (a) Network cost with regards to $I$ & (b) Packet loss rate with regards to $I$ & \\ (c) Patrolling velocity with regards to $I$
\\ \includegraphics[width=3.2in]{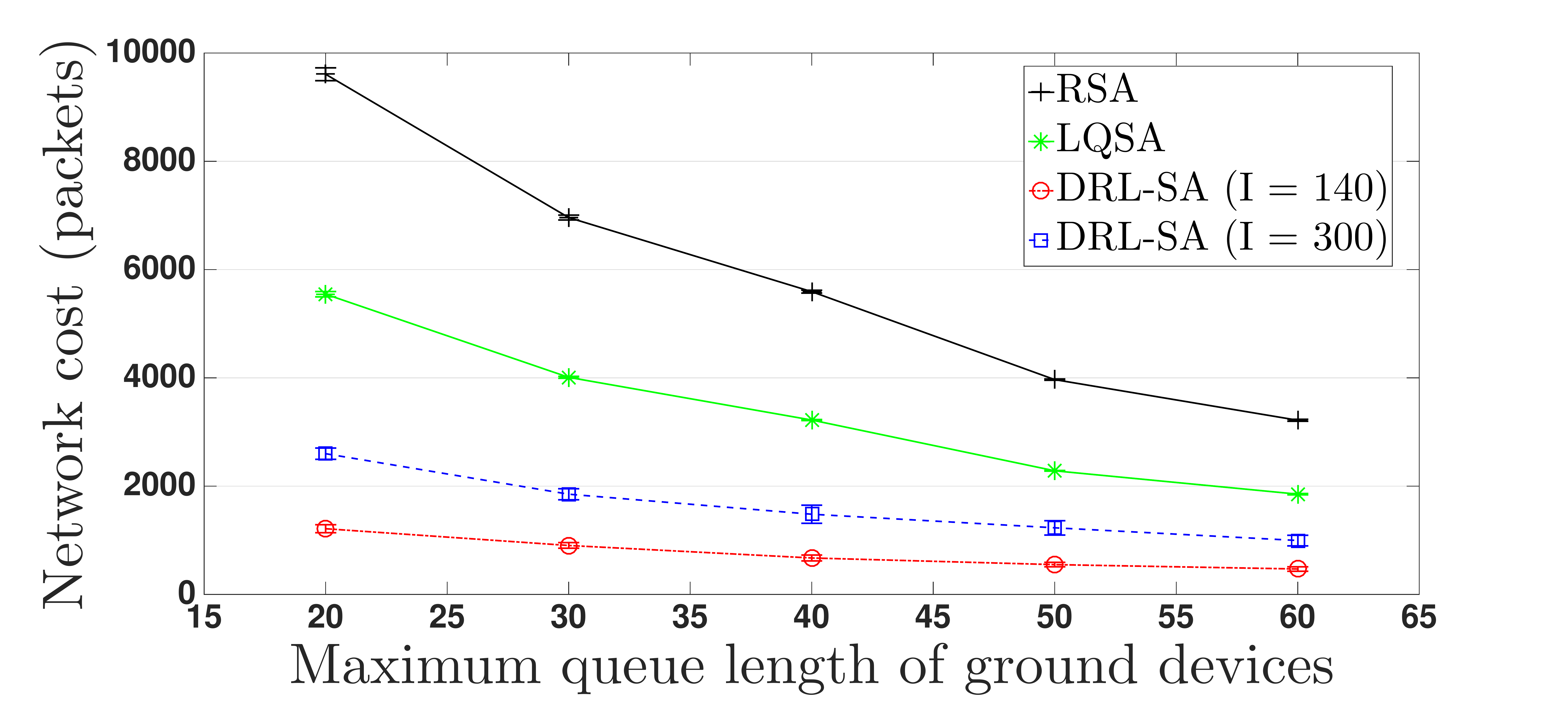}& \includegraphics[width=3.2in]{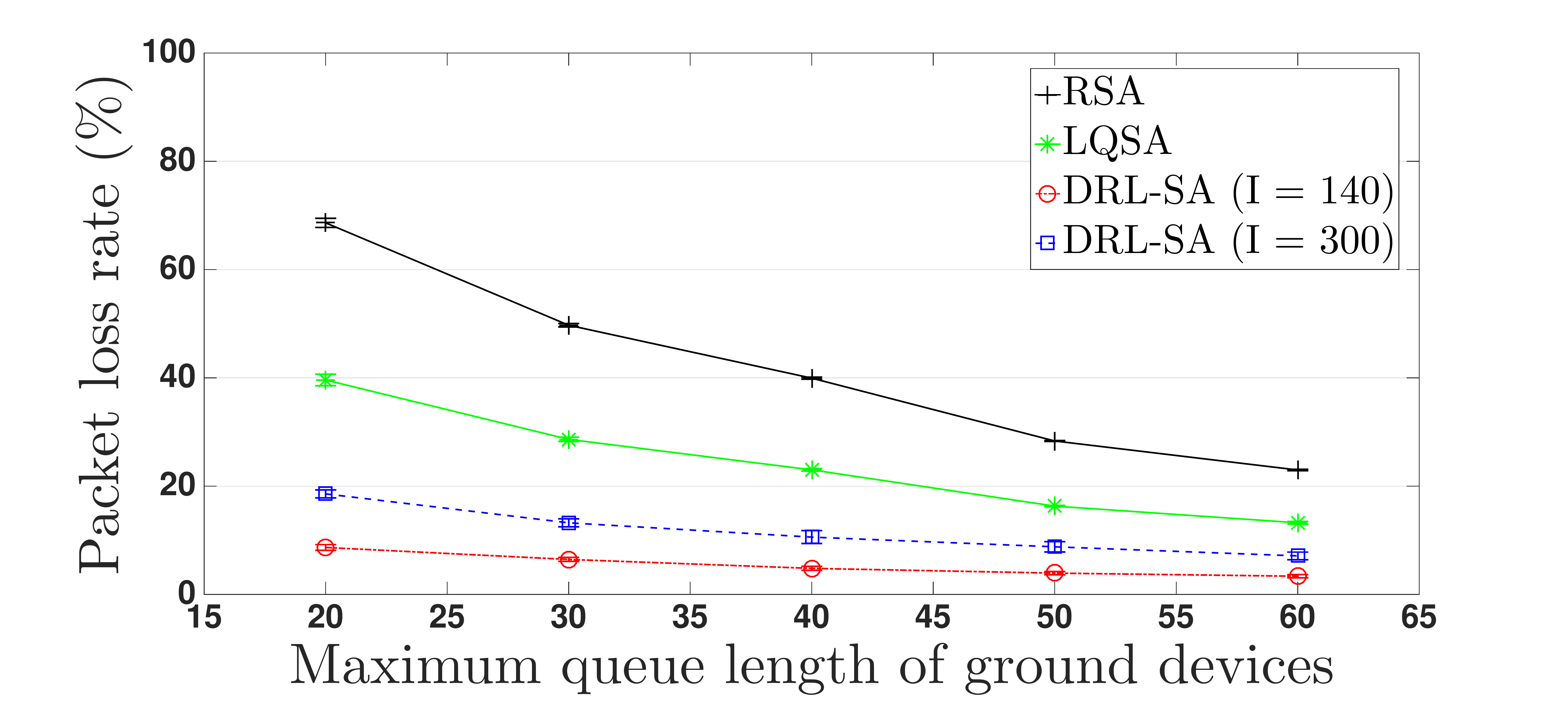}& \\\includegraphics[width=3.2in]{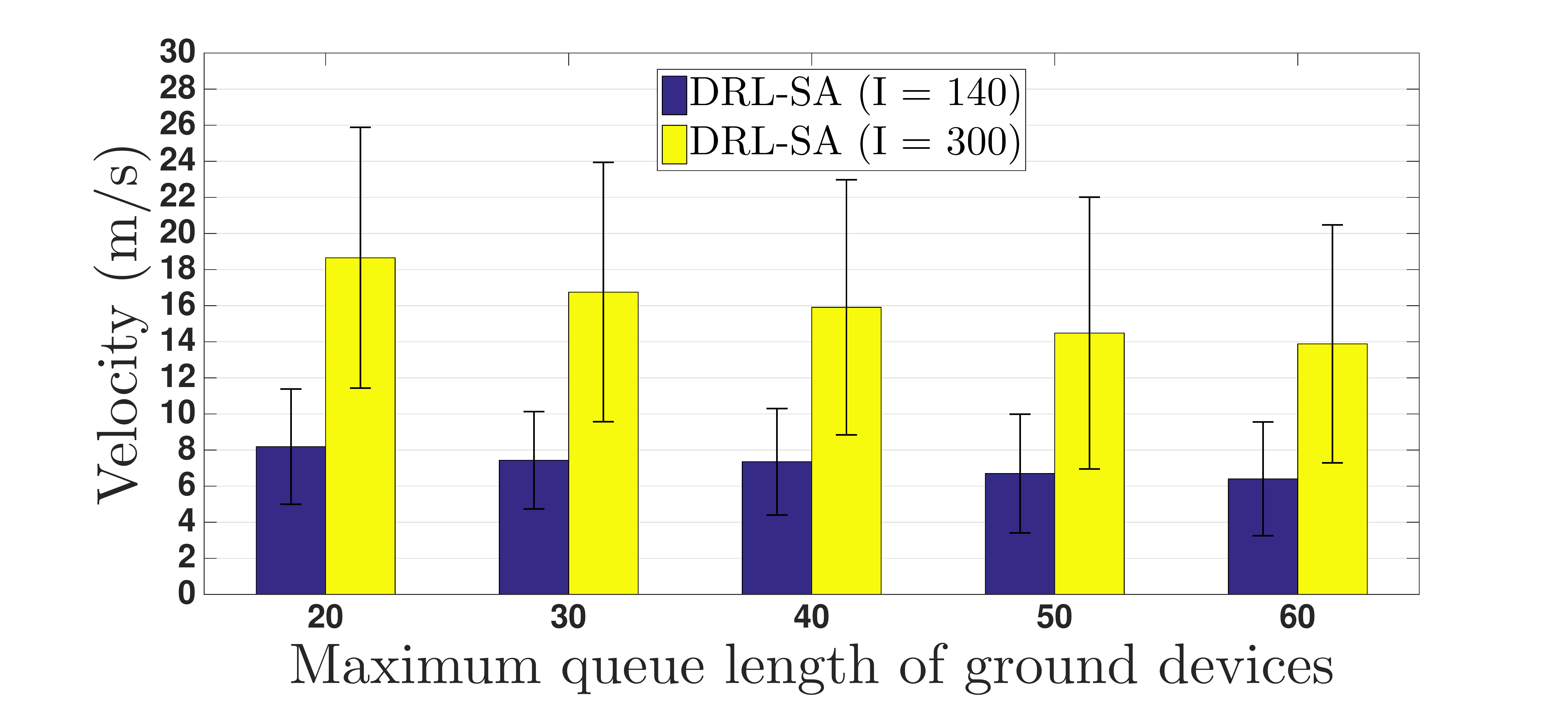} 
\\ (d) Network cost with regards to $D$ & (e) Packet loss rate with regards to $D$ & \\ (f) Patrolling velocity with regards to $D$
\end{tabular}
\end{center}
\caption{A comparison of network cost and packet loss rate by DRL-SA and the typical scheduling strategies, where the error bars show
the standard deviation over 20 runs. The patrolling velocity of the UAV given the different number of ground devices and data queue length is also presented. }
\label{fig_cost_rate_nodes_queue}
\end{figure*}

Figure~\ref{fig_cost_rate_nodes_queue}(a) studies the network cost with an increasing number of ground devices, where the data queue length of DRL-SA is set to 50 or 10. In general, the proposed DRL-SA is able to reduce the network cost to a greater extent than RSA and LQSA.  Particularly, when $I$ = 200 and $D$ = 10, the maximum network cost of DRL-SA is lower than RSA and LQSA by around 82.8\% and 69.2\%, respectively. The performance gains keep growing with $I$. 
The reason is that DRL-SA learns the ground devices' energy consumption and data queue states, so that the scheduling of MPT and data communications can minimize the data packet loss of the entire network. 

Figure~\ref{fig_cost_rate_nodes_queue}(b) illustrates the packet loss rate, which is the ratio of the network cost and the total number of data packets of all the ground devices. Specifically, DRL-SA has a similar packet loss rate to LQSA, when there are 50 devices in the network. However, from $I$ = 80 to 200, DRL-SA outperforms the two non-learning-based algorithms. In particular, when $I$ = 200, DRL-SA with $D = 10$ achieves 53\%, and 25\% lower packet loss rates than RSA and LQSA, respectively. 
In Figure~\ref{fig_cost_rate_nodes_queue}(b), we also see that the packet loss rate of DRL-SA only slightly grows about 2\% from $I$ = 50 to 200. In other words, adding more devices to the network does not result in a critical data packet loss. 
This is because DRL-SA takes the actions to adapt the instantaneous velocity of the UAV to ensure the connection time and channel quality between the device and the UAV, hence minimizing the data packet loss. 
As shown in Figure~\ref{fig_cost_rate_nodes_queue}(c), the patrolling velocity raises with an increase of $I$. This is reasonable because the UAV needs to transfer power and collect data from more devices in order to reduce their data queue overflow. 
Furthermore, it is also observed that increasing the data queue length of the devices downgrades the patrolling velocity of the UAV. The reason is that a larger data queue can hold more packets. This allows an extended data transmission time between the ground device and the UAV and in turn, reduces the patrolling velocity. 

\subsubsection{Data queue length of ground devices}
We consider different data queue lengths of the ground devices, i.e., $D$, where the number of ground devices is set to 140 or 300. 
Figures~\ref{fig_cost_rate_nodes_queue}(d) and~(e) depict the network cost and packet loss rate with respect to the maximum data queue length of the ground devices, respectively. 
We observe that DRL-SA achieves lower network costs and packet loss rates than RSA and LQSA, while DRL-SA outperforms RSA with substantial gains of 82.7\% and 69\% when $D$ = 20 and $I$ = 140. Moreover, given $I = 300$, from $D$ = 20 to $D$ = 60, the network cost and packet loss rate of DRL-SA drop by 62.2\% and 13.2\%. This confirms that DRL-SA significantly reduces data queue overflow for all the ground devices when enlarging their data queue length. In terms of the patrolling velocity of the UAV, it is observed in Figure~\ref{fig_cost_rate_nodes_queue}(f) that DRL-SA reduces the velocity from 18.5 m/s to 14.1 m/s, and from 8.1 m/s to 6.2 m/s, given 300 and 140 ground devices in the network, respectively. 

Figure~\ref{fig_cost_rate_nodes_queue} implies a tradeoff between the data packet loss and the battery lifetime of the UAV. Specifically, an increase of the network size, or a decrease of data queue length, results in a growth of the packet loss due to the data queue overflow. In this case, the patrolling velocity of the UAV can be increased so that more ground devices harvest energy and transmit data, which reduces the packet loss of all the devices. 
However, accelerating the patrolling velocity speeds up draining the energy of the UAV's battery, and reduces the lifetime of the network. 
Therefore, the network size and the data queue length need to be balanced, so as to maintain a sustainable UAV-assisted network while reducing the data packet loss.

\subsubsection{The actions of the UAV}
\begin{figure}[htb]
\centering
\includegraphics[width=4in]{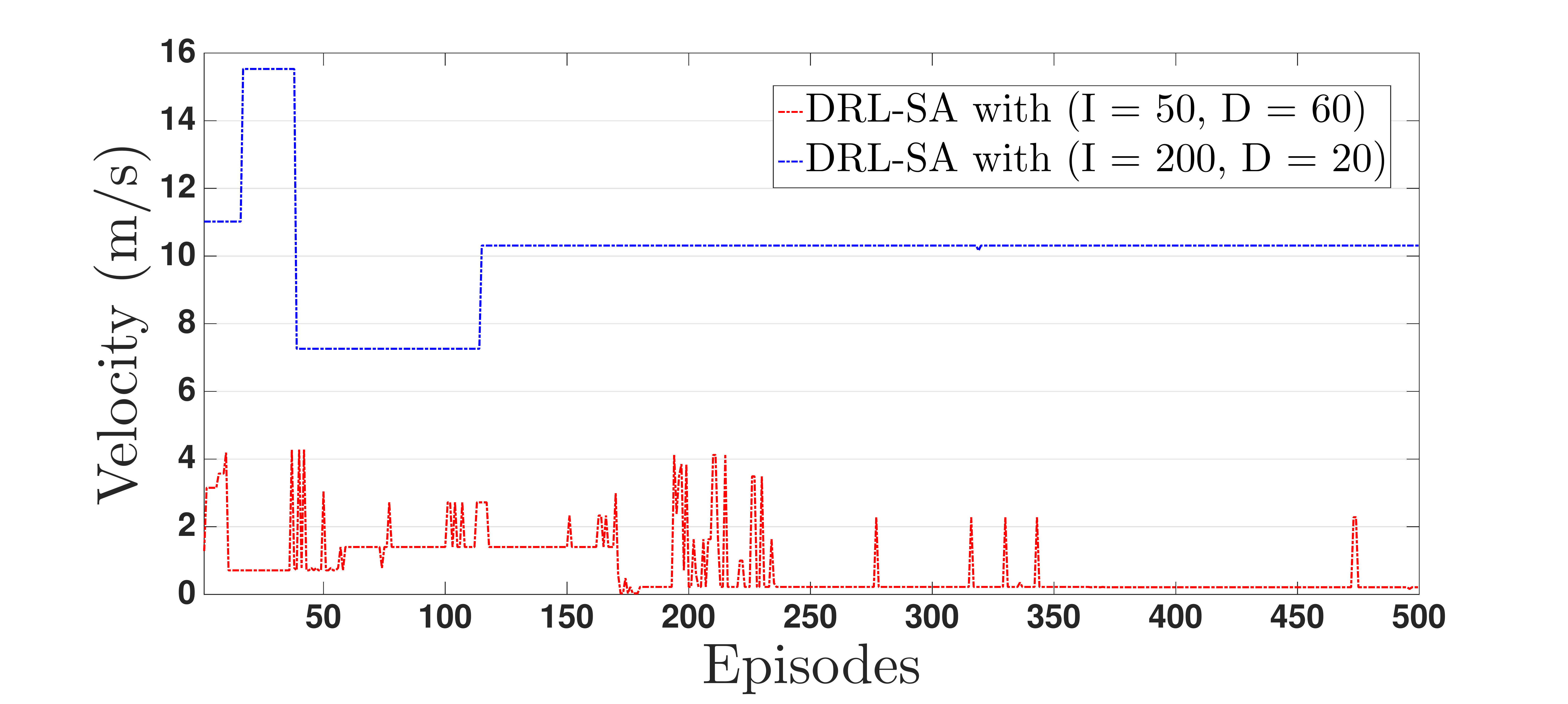}
\caption{\small{The patrolling velocity of the UAV with respect to the episodes, given $(I = 50, D = 60)$ and $(I = 200, D = 20)$.} }
\label{fig_uav_velocity}
\end{figure}

To further reveal the impact of network size and data queue length on the actions of the UAV, Figure~\ref{fig_uav_velocity} demonstrates the patrolling velocity allocated by the proposed DRL-SA, with respect to the episodes, given $(I = 50, D = 60)$ and $(I = 200, D = 20)$. 
We can see that DRL-SA with $(I = 200, D = 20)$ allocates 8.7 m/s higher patrolling velocity to the UAV than the one with $(I = 50, D = 60)$ on average. This confirms that the patrolling velocity drops with a decrease of network size or growth of data queue length, due to the data queue overflow, as explained in Figure~\ref{fig_cost_rate_nodes_queue}. 
Moreover, the result in Figure~\ref{fig_uav_velocity} also presents that DRL-SA allocates the patrolling velocity adapting to the time-varying network state. As observed, the patrolling velocity allocation converges with an increase in the learning time (i.e., episodes). 

\subsubsection{Discount factor of learning}
\begin{figure}[htb]
\centering
\includegraphics[width=4in]{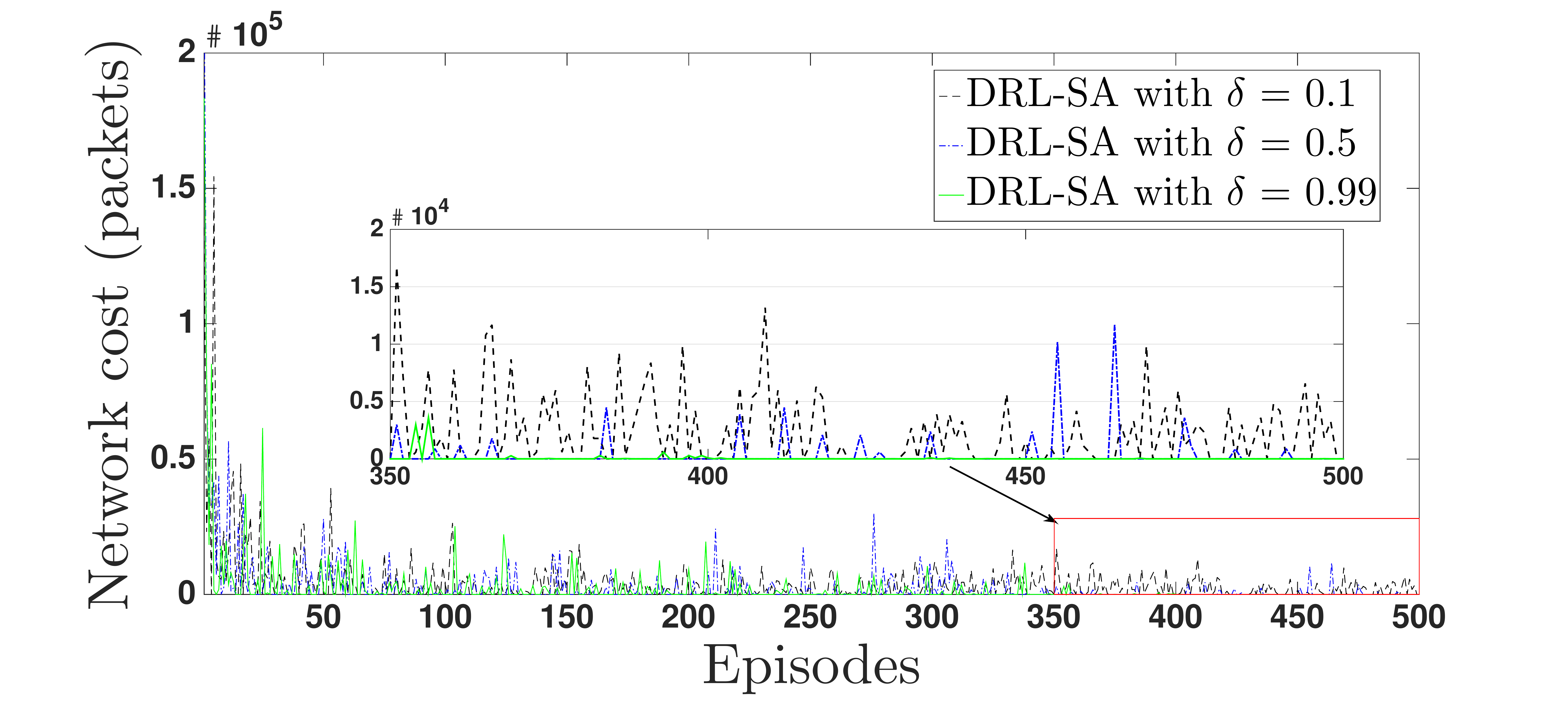}
\caption{\small{Network cost of DRL-SA with regards to the episodes given the discount factor $\delta = 0.1, 0.5$ and $0.99$.} }
\label{fig_cost_discount}
\end{figure}

Since DRL-SA utilizes deep Q learning to approximate the Q function with asymptotic convergence, the convergence time can be affected by the discount factor in the learning process. Figure~\ref{fig_cost_discount} plots the network cost of DRL-SA with regards to the episodes given the discount factor $\delta = 0.1, 0.5$ and $0.99$. $I$ and $D$ are set to 300 ground devices and 20, respectively. The other settings are provided in Section~\ref{sec_config}. 

As observed, DRL-SA has a high network cost at the beginning of the learning. The performance improves with the increase of the episodes due to deep reinforcement learning. 
In particular, the network cost of DRL-SA with $\delta$ = 0.99 quickly falls to 0 from episode 1 to episode 359, and the performance remains relatively stable afterward. However, the convergence of DRL-SA with $\delta$ = 0.5 or 0.1 requires more than 475 or 500 episodes. 
Therefore, the result in Figure~\ref{fig_cost_discount} indicates that the convergence rate of DRL-SA grows with $\delta$. In other words, a high discount factor of learning accelerates the learning process of the deep Q-network, as confirmed by~\eqref{eq_optQfunc}. 

%=============================================================================%
%============================Section 7 Conclusion=================================%
\section{Conclusion}
\label{sec_cond}
In this paper, we focus on online MPT and data collection in the presence of on-board control of the patrolling velocity of the UAV, for preventing battery drainage and data queue overflow of the sensing devices. 
The problem is formulated as the MDP, with the states of battery level and data queue length of the ground devices, channel conditions, and waypoints given the trajectory of the UAV. 
We propose an on-board deep Q-network that can enlarge the state and action space of the MDP to minimize the data packet loss of the entire system. 
Based on deep reinforcement learning, DRL-SA is developed to learn the optimal resource allocation strategy asymptotically through online training at the on-board deep Q-network, where the selection of the ground device, modulation scheme, and instantaneous patrolling velocity of the UAV are jointly optimized. 
Moreover, DRL-SA carries out experience replay to reduce expansion of the state space in which the algorithm's scheduling experiences at each time step are stored in a data set. 

The proposed DRL-SA scheme is implemented by using Keras deep learning library with Google TensorFlow as the backend engine. Numerical results demonstrate that DRL-SA reduces packet loss by at least 69.2\%, as compared to the existing non-learning greedy algorithms.

\section*{Acknowledgements}
This work was partially supported by National Funds through FCT/MCTES (Portuguese Foundation for Science and Technology), within the CISTER Research Unit (CEC/04234); also by the Operational Competitiveness Programme and Internationalization (COMPETE 2020) through the European Regional Development Fund (ERDF) and by national funds through the FCT, within project POCI-01-0145-FEDER-029074 (ARNET). 

\appendix
\section{[Optimizing $\phi_i^z(t)$]}
To minimize the packet loss stemming from insufficient energy, $\phi_i^z(t)$ of device $i$ is to be chosen to maximize the energy harvested during a contact time with the UAV, with a length of $\widehat{T}^z_i(t)$. 
The optimal modulation of the ground device, $\phi^z_i(t)^{\star}$, is independent of the battery level and the queue length. This is because $\phi^z_i(t)^{\star}$ is selected to maximize the increase of the battery level at device $i$, under the bit error rate requirement $\epsilon_i$ for the packet transmitted. As a result, $\phi_i^z(t)$ can be decoupled from $\mathcal{A}$, and optimized in prior by
\begin{align}
\phi_i^z(t)=&\arg\max_{\phi=1,\cdots,\Phi} \bigg\{(\widehat{T}^z_i(t)-\frac{B}{\phi W}) P_i^z(t) -\frac{B}{\phi W}P_i^z(\phi)\bigg\},
\label{eq_rho}
\end{align}
the right-hand side (RHS) of which, by substituting \eqref{eq_powerTx} and \eqref{eq_txPower}, can be rewritten as
\begin{align}
\max_{\phi=1,\cdots,\Phi} \bigg\{(\widehat{T}^z_i(t)-\frac{B}{\phi W}) &\omega(d_i^z(t),\theta_i^z(t)) P_{\rm UAV}^{tx} \|\mathbf{h}_i^z(t)\|^2 - \nonumber \\
&\frac{B\kappa_2^{-1}\ln(\frac{\kappa_1}{\epsilon})}{\|\mathbf{h}_i^z(t)\|^2\phi W}\big(2^{\phi}-1\big)\bigg\}, 
\label{eq_maxRho}
\end{align}
where $W$ is the bandwidth of the uplink data transmission, $\frac{1}{W}$ is the duration of an uplink symbol, and $\frac{B}{\phi W}$ is the duration of uplink data transmission. $(\widehat{T}^z_i(t)-\frac{B}{\phi W})$ is the rest of the time slots used for downlink WPT, and $\widehat{T}^z_i(t)$ is the contact time between the ground device and the UAV in the time slot, which is affected by the patrolling velocity of the UAV. Thus, we have
\begin{align}
\widehat{T}^z_i(t) = \frac{2\sqrt{(d_i^z(t))^2-(b^z)^2}}{v^z(t)}, 
\label{eq_duration}
\end{align}
where $b^z$ is the altitude of the UAV at lap $z$. We assume that the UAV maintains the same altitude and the same heading in each lap.

By using the first-order necessary condition of the optimal solution, we have 
\begin{align}
\frac{d}{d\phi}((\widehat{T}^z_i(t)-\frac{B}{\phi W}) &\omega(d_i^z(t),\theta_i^z(t)) P_{\rm UAV}^{tx} \|\mathbf{h}_i^z(t)\|^2- \nonumber \\
&\frac{B\kappa_2^{-1}\ln(\frac{\kappa_1}{\epsilon})}{\|\mathbf{h}_i^z(t)\|^2\phi W}\big(2^{\phi}-1\big)) = 0, 
\end{align}

\begin{align}
&\phi^{-2}\frac{B}{W} \omega(d_i^z(t),\theta_i^z(t)) P_{\rm UAV}^{tx} \|\mathbf{h}_i^z(t)\|^2 - \frac{B\kappa_2^{-1}\ln(\frac{\kappa_1}{\epsilon})}{\|\mathbf{h}_i^z(t)\|^2 W} \cdot \nonumber \\
&~~~~(\phi^{-1}2^{\phi}\text{In}2-\phi^{-2}2^{\phi}) - \frac{B\kappa_2^{-1}\ln(\frac{\kappa_1}{\epsilon})}{\|\mathbf{h}_i^z(t)\|^2 W}\phi^{-2} = 0. 
\end{align}

The $\phi$ values are then given as follows: 
\begin{align}
&\phi2^{\phi}\text{In}2-2^{\phi} = \nonumber \\
&\frac{B}{W} \omega(d_i^z(t),\theta_i^z(t)) P_{\rm UAV}^{tx} \|\mathbf{h}_i^z(t)\|^2 \frac{\|\mathbf{h}_i^z(t)\|^2 W}{B\kappa_2^{-1}\ln(\frac{\kappa_1}{\epsilon})} - 1. 
\label{eq_rhofinal}
\end{align}
Since the left-hand side (LHS) of~\eqref{eq_rhofinal} monotonically increases with $\phi$, the optimal value $\phi^{\star}$ can be obtained by applying a bisection search method, and evaluating the two closest integers about the fixed point of the bisection method~\cite{estep2002bisection}. 
Specifically, $\phi_- = 1$ and $\phi_+ = \Phi$ are initialized. Each iteration of the bisection method contains 4 steps applied over the range of $\phi = [1, \Phi]$, as follows. 
\begin{itemize}
\item The midpoint of the modulation interval $[\phi_-;\phi_+]$ is calculated, which gives $\phi_{\rm mid} = \frac{\phi_-+\phi_+}{2}$. 
\item Substitute $\phi_{\rm mid}$ into~\eqref{eq_rhofinal} to obtain the function value $f(\phi_{\rm mid})$. 
\item If the convergence is attained (that is, the modulation interval or $|f(\phi_{\rm mid})|$ cannot be further reduced), return $\phi_{\rm mid}$ and stop the iteration.
\item Replace either $(\phi_-, f(\phi_-))$ or $(\phi_+, f(\phi_+))$ with $(\phi_{\rm mid}, f(\phi_{\rm mid}))$. 
\end{itemize}

\bibliographystyle{elsarticle-num}
\bibliography{EHUAV}  

\end{document}